# Material properties of biomolecular condensates emerge from nanoscale dynamics

Nicola Galvanetto[1,2,*], Miloš T. Ivanović[1,*] , Simone A. Del Grosso[1], Aritra Chowdhury[1], Andrea Sottini[1], Daniel Nettels[1], Robert B. Best[3,*] and Benjamin Schuler[1,2,*]

[1]Department of Biochemistry, University of Zurich, Zurich, Switzerland
[2]Department of Physics, University of Zurich, Zurich, Switzerland
[3]Laboratory of Chemical Physics, National Institute of Diabetes and Digestive and Kidney Diseases, National Institutes of Health, Bethesda, MD, USA
* Corresponding authors: N. Galvanetto (n.galvanetto@bioc.uzh.ch), M. T. Ivanović (m.ivanovic@bioc.uzh.ch), R. B. Best (robert.best2@nih.gov), B. Schuler (schuler@bioc.uzh.ch)

**This article is published in *PNAS*: https://doi.org/10.1073/pnas.2424135122**

## Abstract

Biomolecular condensates form by phase separation of biological polymers and have important functions in the cell — functions that are inherently linked to their physical properties at different scales. A notable aspect of such membraneless organelles is that their viscoelastic properties can vary by orders of magnitude, but it has remained unclear how these pronounced differences are rooted in the nanoscale dynamics at the molecular level. Here we investigate a series of condensates formed by complex coacervation of highly charged disordered proteins and polypeptides that span about two orders of magnitude in bulk viscosity. We find that their viscosity is highly correlated with protein translational diffusion and nano- to microsecond chain dynamics. Remarkably, analytical relations from polymer physics can predict condensate viscosity from diffusivity and chain dynamics, and vice versa, even for more hydrophobic disordered proteins and for synthetic polyelectrolytes, indicating a mechanistic link across several decades of length- and timescales. Atomistic simulations reveal that the observed differences in friction — a key quantity underlying these relations — reflect differences in inter-residue contact lifetimes as a function of arginine content and salt concentration, leading to the vastly different dynamics among condensates. The rapid exchange of inter-residue contacts we observe may be a general mechanism for preventing dynamic arrest in compartments densely packed with polyelectrolytes, such as the cell nucleus.

## Significance Statement

Cells organize many of their activities within membraneless compartments known as biomolecular condensates. Our research uncovers how the rapid, small-scale motions of disordered proteins within these condensates determine their overall physical characteristics, such as viscosity and molecular transport. Remarkably, we found accurate relationships between nanoscale protein dynamics and the mesoscale behavior of condensates formed by charged, intrinsically disordered proteins. They provide quantitative predictions connecting independently measurable quantities at different scales. This finding bridges a gap in our understanding of the multi-scale organization of the cell, and opens up possibilities for understanding and quantitatively influencing the properties of these essential cellular structures based on their molecular interactions.

## Main Text

### Introduction

A substantial fraction of all cellular proteins are organized in biomolecular condensates (1) formed as a consequence of phase separation, an intriguing feature of subcellular organization (2–4). These membraneless bodies can regulate cellular homeostasis and coordinate numerous biological functions through the assembly of proteins and nucleic acids (5–7). The underlying cellular processes span a wide spectrum of time- and length-scales (8), and they are governed by the physical properties of the condensates (9) and the molecular driving forces that lead to phase separation (10–14). At the nanoscale, the rate at which biomolecules explore different conformations determines the efficiency of biochemical interactions and reactions (15–17). At the microscale, these processes and their spatial organization are controlled by the translational diffusion of biomolecules within phase-separated organelles as well as the biomolecular exchange with the environment (18, 19). At the mesoscale, material properties play a role; for example, bulk viscosity affects the fusion times of condensates (20, 21), which can influence cell fate (22, 23). This multi-scale complexity poses a considerable challenge in deciphering the relationships between these dynamic processes and in establishing quantitative relations that can predict and explain the behavior of biomolecular condensates. The nanoscale dynamics of biomolecular conformations are expected to be related to translational diffusion (24) and to the emergent bulk viscosity of the percolated network they form (25). Material properties ultimately derive from the interaction strengths among the biomolecules that drive phase separation, and therefore from their specific amino acid sequences (26–33), but how molecular and mesoscale dynamics are linked quantitatively is an open question (34).

A biological example with this multi-scale complexity is the cell nucleus (7, 35, 36) which is rich in highly charged biomolecules. To compensate for the high negative net charge of DNA, highly positively charged proteins, such as histones and protamines, have evolved to compact the chromosomes (37, 38). Other highly charged intrinsically disordered proteins (IDPs) are involved in remodeling chromatin and in regulating gene expression and replication. For instance, the negatively charged prothymosin α (ProTα) can sequester histone H1 and accelerate its dissociation from nucleosomes (39, 40). The two oppositely charged disordered proteins histone H1 and ProTα form viscous droplets by complex coacervation at near-physiological salt concentrations, but maintain surprisingly rapid dynamics at the molecular level (41). However, viscosities and chain dynamics are expected to depend on the composition of these biological polyelectrolytes as well as the solution conditions, especially the salt concentration. Here we aim to identify general relations between the molecular and mesoscopic properties of biomolecular condensates across a wide range of dynamics.

We focus on complex coacervates between highly charged intrinsically disordered proteins and peptides. In the condensates they form, associative phase separation is driven by electrostatic interactions (42–45) and is thus highly sensitive to salt concentration and the identity of the charged residues (46, 47). We employ a combination of single-molecule techniques to investigate the conformational and translational dynamics of the polypeptides, and microrheology to assess mesoscale properties. We find that the chain dynamics of intrinsically disordered proteins within these condensates range from hundreds of nanoseconds to tens of microseconds. These reconfiguration times correlate linearly with the translational diffusion coefficients of the proteins and the bulk viscosity of the condensates. From large-scale all-atom molecular dynamics (MD) simulations, we find that low salt concentrations and especially the presence of arginine residues increase the lifetimes of inter-chain contacts, which slows down larger-scale condensate dynamics. Altogether, we thus demonstrate a direct relation between the nanoscopic dynamics of protein chain reconfiguration, microscale translational motion, and mesoscopic viscosity within biomolecular condensates. These relations can be rationalized within the framework of semidilute polymer solutions and generalized to predict the behavior of other condensates across scales.

## Results

### Phase separation of biological polyelectrolytes

To be able to assess the influence of amino acid sequence and composition, we used the highly negatively charged disordered protein ProTα in combination with four positively charged IDPs and polypeptides with different charge densities and amino acid composition (Figure 1A): the lysine-rich histone H1 (net charge +53), the arginine-rich protamine (net charge +22), and two disordered homopolypeptides with 50 lysine (K50) or arginine (R50) residues, respectively (both net charge +50). The strong electrostatic interactions between ProTα and each of the four positively charged partners leads to associative phase separation when mixed at charge-balanced stoichiometries (Figure S1), as expected for oppositely charged polyelectrolytes (44, 45). Viewed under a light microscope, all phase-separated samples appear visually indistinguishable, with an aqueous dilute phase and spherical droplets of dense phase (Figure 1B-D). For all of them, the dense phase has a total protein mass concentration above 100 mg/mL. However, phase separation of the four samples responds differently to salt concentration: At an initial protein concentration of ~10 μM, phase separation is difficult to achieve above 200 mM KCl for the lysine-rich polypeptides, but the arginine-rich polypeptides readily phase-separate with ProTα at higher salt — for R50 even above 1 M KCl. This observation and the corresponding phase diagrams (Figure 1E) highlight quantitative differences in the nature of the interactions of these two positively charged residues (47–51) related to differences in their chemical structure, charge distribution, and polarizability (52). But how do these different interactions affect the conformations and dynamics of the polypeptides that make up the condensates, as well as the corresponding mesoscopic properties?

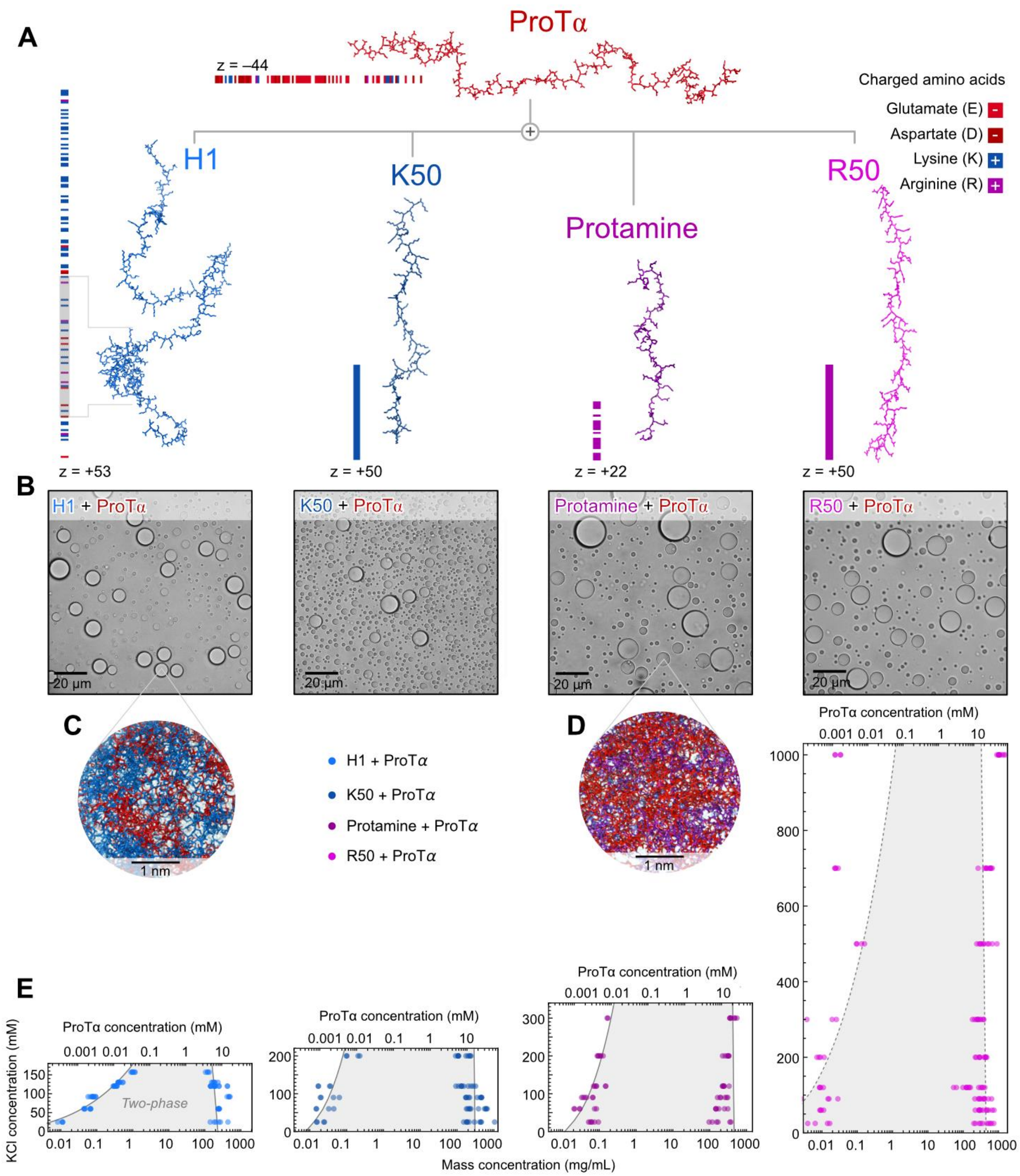

***Figure 1. Phase separation of charged polypeptides strongly depends on their amino acid sequences.*** *(**A**) Illustration of polymers used in this study with the distributions of charges along their sequences, and net charges (z): ProTα, protamine, and H1 are naturally occurring polycationic proteins; poly-L-arginine-50 (R50) and poly-L-lysine-50 (K50) are synthetic polycations (Supplementary Table 1). The gray band in the H1 sequence indicates the globular domain. (**B**) Brightfield microscopy images of phase-separated samples of ProTα mixed with a polycation (protamine, H1, R50, or K50) at charge balance in TEK buffer at 90 mM KCl (ionic strength 98 mM). Scale bar, 20 µm. (**C**) Illustration of the polymer networks on the nanoscale in the dense phases of H1 + ProTα and (**D**) protamine + ProTα from molecular dynamics simulations. (**E**) Phase diagrams from coexistence measurements of dense and dilute phases as a function of salt concentration.*

*The total mass concentration of both components (bottom axis) is based on the measured ProTα concentrations (top axis) and the charge-balanced ratio at which ProTα and the positive partner were mixed* (41) *(ProTα:H1 1.2:1, ProTα:K50 1.13:1, ProTα:protamine 0.5:1, ProTα:R50 1.13:1, see Figure S1). Phenomenological fit with a binodal curve based on Voorn–Overbeek theory* (53) *(solid line, dashed for ProTα:R50 where the theory fails to capture the complex interactions of arginine beyond electrostatics).*

**Condensate dynamics across scales**

To probe the conformations and intrachain dynamics of individual proteins within the different dense phases at the nanoscale, we used confocal single-molecule Förster resonance energy transfer (FRET) spectroscopy (54). We prepared droplets with unlabeled samples and doped them with ProTα double-labeled with Cy3B as a donor and CF660R as an acceptor fluorophore at positions 56 and 110 to monitor intramolecular distances and distance fluctuations. The doping ratio between labeled and unlabeled protein was adjusted to yield a final concentration of ~100 pM labeled ProTα within the droplets to enable FRET measurements with single-molecule resolution (Figure 2A, B). The resulting FRET efficiency histograms (Figure 2C) show that free monomeric ProTα in dilute solution is expanded at low salt concentration, resulting in a low mean transfer efficiency, $\langle E \rangle$, due to the repulsion between the negative charges along the chain. The repulsion is screened at high salt, leading to chain compaction (55, 56) (Figure 2D). The higher FRET efficiencies of ProTα inside the condensate droplets indicate chain compaction, which increases with the charge density and the arginine content of the polycationic interaction partners, resulting in stronger interactions with ProTα (Figure 2C). In contrast to the free monomeric chain, ProTα within the droplets experiences a slight expansion with increasing salt concentration, as indicated by a decrease in $\langle E \rangle$ (Figure 2D). Since we observe no significant correlation between protein mass concentration and chain dimensions (Figure S2A), the most likely cause of this expansion is the screening of the electrostatic attraction between oppositely charged chains by salt.

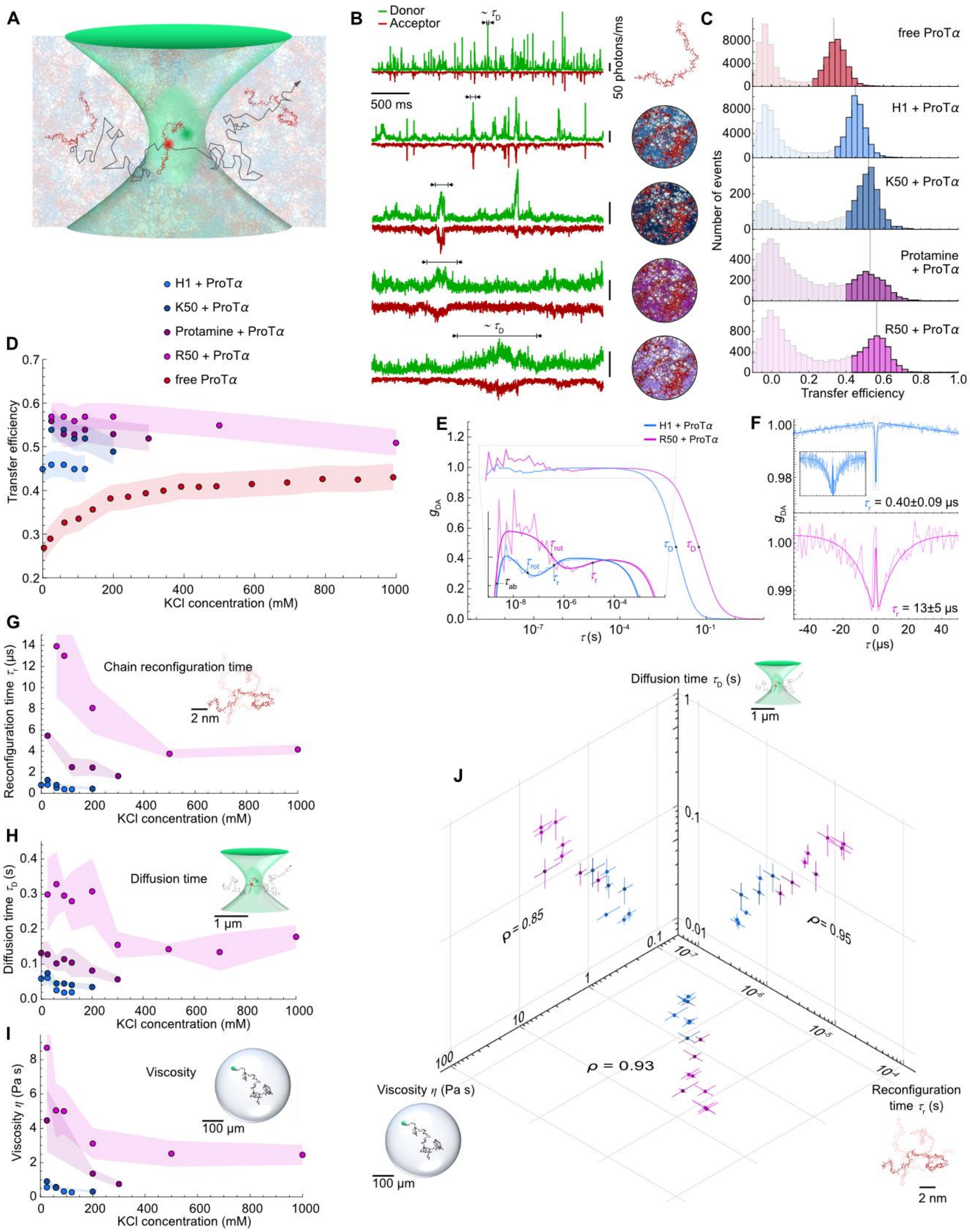

***Figure 2. Single-molecule spectroscopy and microrheology in phase-separated droplets.*** *(**A**) Illustration of a double-labeled ProTα molecule in the dense phase diffusing through the confocal volume. (**B**) Fluorescence time traces of (from top to bottom) double-labeled ProTα as a monomer free in solution, in complex coacervate droplets of ProTα + H1, ProTα + K50, ProTα + protamine, and ProTα + R50. The diffusion time, $\tau_D$, is the average time it takes a single labeled ProTα molecule to transit the confocal volume, resulting in a fluorescence burst. (**C**) Single-molecule*

*transfer efficiency histograms of double-labeled ProTα as a monomer in solution and within droplets (ordered as in **B**) in TEK buffer at 90 mM KCl (ionic strength 98 mM). To minimize the contribution of donor-only molecules and the influence of photobleaching, fluorescence bursts corresponding to the shaded parts of the histograms were excluded from correlation analysis. (**D**) Average transfer efficiency of double-labeled ProTα as a monomer free in solution and in complex coacervate droplets as a function of salt concentration. Shaded bands represent the systematic uncertainty due to instrument calibration. (**E**) Full FCS curves with logarithmic time binning of donor-acceptor cross-correlations ($g_{DA}$) normalized to an amplitude of 1 at 10 μs for ProTα + H1 and 100 μs for ProTα + R50, respectively, to facilitate direct comparison ($\tau_{rot}$, segmental rotational correlation time; $\tau_r$, chain reconfiguration time; $\tau_D$, translational diffusion time). (**F**) Representation of FCS curves with linear time binning in the range where chain dynamics dominate the signal. (**G**) ProTα reconfiguration time, $\tau_r$, in the different coacervates as a function of salt concentration obtained from the FCS fits as shown in **E/F** (see Methods). Error bands, standard deviations from three measurements or the error of the fit of $\tau_r$, whichever was greater (see Methods). (**H**) Translational diffusion time through the confocal volume of double-labeled ProTα in the different coacervates as a function of salt concentration obtained from the FCS fits as shown in **E** (see Methods). Error bands, standard deviations from $n \geq 3$ measurements. (**I**) Viscosity, η, from measurements of translational diffusion of 100- and 500-nm polystyrene beads within the different coacervates from particle tracking (see Methods) as a function of salt concentration. Error bands, standard deviation from $n \geq 20$ tacked beads. (**J**) Correlations between the data in **G**, **H** and **I** indicate a physical relation between $\tau_r$, $\tau_D$, and η. The slopes from linear regression of the data, plotted on a log scale, are: 1.3 ± 0.2 for $\tau_D$ vs $\tau_r$; 1.2 ± 0.2 for η vs $\tau_r$; 0.8 ± 0.2 for η vs $\tau_D$ (uncertainties represent 95% confidence intervals, ρ are the Pearson correlation coefficients).*

The intrinsically disordered protein ProTα samples a heterogeneous ensemble of conformations within the droplets (41). We measured the corresponding chain relaxation (25) or reconfiguration times, $\tau_r$, by monitoring the fluctuations of the acceptor-donor distance using single-molecule FRET combined with nanosecond fluorescence correlation spectroscopy (nsFCS) (57, 58) (Figure 2E, F, see Methods). We find that the identity of the polycationic partner strongly influences the chain dynamics of ProTα in the dense phases. ProTα and the lysine-rich H1 form droplets in which the protein rearrangements are extremely fast, with $\tau_r$ of hundreds of nanoseconds (41), whereas in arginine-rich droplets, chain reconfiguration is slowed down by more than an order of magnitude, with $\tau_r$ exceeding 10 μs in some cases (Figure 2F, G). In addition to the dependence on sequence composition, $\tau_r$ in the droplets increases with decreasing salt concentration, by a factor of about 2 to 3 over the salt concentrations accessible for the different condensates (Figure 2G). This observation is consistent with the hypothesis that ions screen the intermolecular interactions that slow down chain rearrangements, as reflected by the moderate chain expansion at high salt concentration (Figure 2D).

From the FCS measurements (Figure 2E), we can also extract the diffusion time, $\tau_D$, of the labeled protein molecules through the confocal volume (Figure 2H, see Methods). While $\tau_r$ reports on the nanoscopic dynamics within the polypeptide chain, $\tau_D$ provides information on the translational motion of the protein through the percolated network of the condensate on the micrometer length scale of the confocal volume and is inversely proportional to the diffusion coefficient. The dependence of the translational diffusion of ProTα on the sequence composition of the binding partner and the salt concentration shows remarkably similar trends as the nanoscopic chain dynamics (compare Figure 2H and G): ProTα diffuses more rapidly in droplets with lysine-rich than with arginine-rich interaction partners and at high than at low salt concentrations.

To characterize the mesoscopic dynamics of the condensates, we used microrheology and tracked the diffusion of fluorescent beads of 100 and 500 nm diameter inside the droplets (Figure S3A, B). From the mean squared displacement, we obtained the viscosity from the Stokes–Einstein relation (Methods). Viscosity is a length-scale dependent quantity in condensates (41, 59), but in this study we focus on the bulk viscosity by using beads much larger than the correlation lengths of the protein networks (25) (Methods). The viscosity in the droplets is remarkably different for complex coacervates with different polycationic proteins and ranges from ~300 to ~10,000 times the viscosity of water (Figure 2I). These values remain constant for days in a given sample, indicating the absence of aging effects (60) over this period.

In summary, the salt concentration and especially the amino acid sequence composition have a strong influence on the dynamic properties of the condensates across length- and timescales, from the nanoscopic chain reconfiguration time and the microscopic translational diffusion time of molecules to the viscosity at the mesoscopic scale of entire droplets (Figure 2G–I). The changes span nearly two orders of magnitude for each of the physical properties studied, with remarkably high correlations between them (Figure 2J), suggesting an underlying causal link across scales. To identify the molecular origins of the experimentally observed behavior, we turned to large-scale molecular dynamics (MD) simulations.

**Interaction dynamics from atomistic simulations**

To be able to probe interactions at the atomic level and compare absolute timescales with experiment, we used all-atom MD simulations of the coacervates with explicit solvent in a recently validated (41) slab configuration (61) (see Methods). To assess the role of lysine versus arginine, we simulated systems consisting of 96 ProTα and 80 H1 molecules in one case, and 96 ProTα and 197 protamine molecules in the other. The two systems correspond to roughly 4 and 2.6 million atoms in the simulation box, respectively (Figure 3A, B, Videos S1 and S2). To study the effect of salt concentration, we performed simulations with 8 mM and 128 mM KCl for both systems.

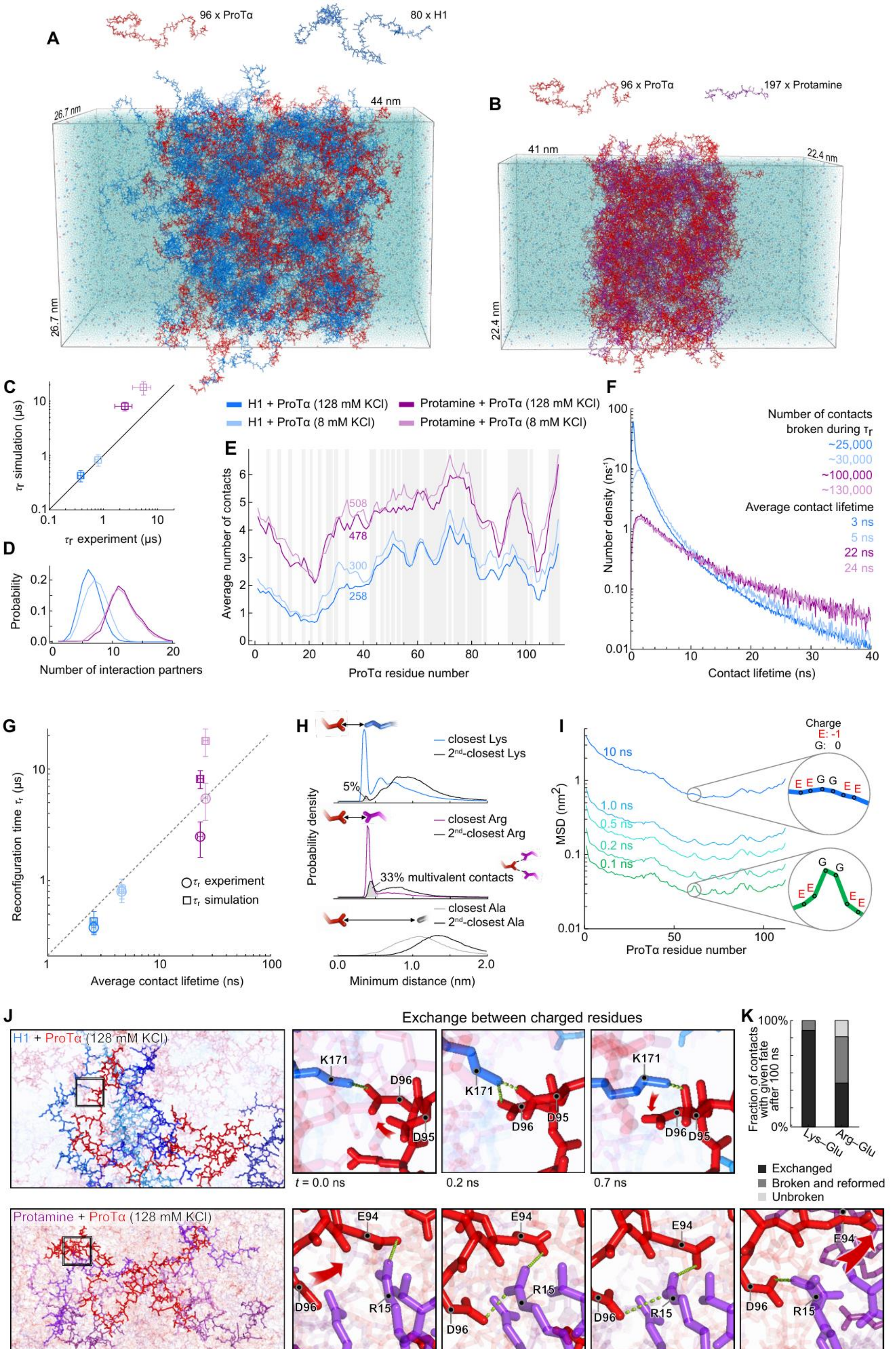

A
96 x ProTα
80 x H1
26.7 nm
44 nm
26.7 nm
B
96 x ProTα
197 x Protamine
41 nm
22.4 nm
22.4 nm
C
τr simulation (μs)
τr experiment (μs)
D
Probability
Number of interaction partners
H1 + ProTα (128 mM KCl)
H1 + ProTα (8 mM KCl)
Protamine + ProTα (128 mM KCl)
Protamine + ProTα (8 mM KCl)
E
Average number of contacts
ProTα residue number
508
478
300
258
F
Number of contacts broken during τr
~25,000
~30,000
~100,000
~130,000
Average contact lifetime
3 ns
5 ns
22 ns
24 ns
Number density (ns-1)
Contact lifetime (ns)
G
Reconfiguration time τr (μs)
Average contact lifetime (ns)
τr experiment
τr simulation
H
closest Lys
2nd-closest Lys
5%
closest Arg
2nd-closest Arg
33% multivalent contacts
closest Ala
2nd-closest Ala
Probability density
Minimum distance (nm)
I
Charge
E: -1
G: 0
10 ns
1.0 ns
0.5 ns
0.2 ns
0.1 ns
MSD (nm2)
ProTα residue number
J
Exchange between charged residues
H1 + ProTα (128 mM KCl)
K171
D96
D95
t = 0.0 ns
0.2 ns
0.7 ns
Protamine + ProTα (128 mM KCl)
E94
R15
D96
t = 0 ns
6 ns
10 ns
16 ns
K
Fraction of contacts with given fate after 100 ns
Lys–Glu
Arg–Glu
Exchanged
Broken and reformed
Unbroken

***Figure 3. All-atom simulations of dense phases at different salt concentrations.***
*(**A**) All-atom explicit-solvent simulations of 96 ProTα (red) and 80 H1 molecules (blue) and (**B**) of 96 ProTα (red) and 197 protamine molecules (purple) in slab geometry* (41, 61)*, including water (transparent blue spheres), $K^+$ ions (blue spheres), and $Cl^-$ ions (red spheres). (**C**) Comparison between the experimental chain reconfiguration time, $\tau_r$, and the corresponding distance decorrelation time between residues 58 and 112 (corresponding to the dye positions in the experiments) from simulations ($\tau_r$ of protamine–ProTα slab at 8 mM KCl concentration is compared with the value measured at 25 mM KCl due to experimental limitations in performing stable single-molecule recordings at lower salt conditions; for uncertainties, see Methods). (**D**) Distribution of the number of H1 and protamine molecules simultaneously in contact with a single ProTα. (**E**) Average number of contacts (Cα atoms closer than 1 nm) made by each residue of ProTα in the four dense phases, with the average total number of contacts indicated. Gray bands indicate negatively charged residues. (**F**) Distribution of the lifetimes of contacts made by ProTα residues in the four dense phases (see Methods for detailed contact lifetime definition). We also report the estimated number contacts of a single ProTα chain that break during its reconfiguration time. (**G**) The correlation between the average contact lifetime of individual ProTα residues with residues in other chains (averages over all residues of all ProTα molecules in each simulation) and the chain reconfiguration time suggests a frictional effect of intermolecular contacts slowing down chain dynamics (for uncertainties, see Methods). (**H**) Distance distribution of the closest and the second-closest lysine (charge +1), arginine (+1), and alanine (0) to glutamate (-1) residues in ProTα chains (lysine and alanine distributions from H1–ProTα simulation, arginine distributions from protamine–ProTα simulation, both at 128 mM KCl). A sharp peak is present only in the distributions between oppositely charged residues. The shaded gray area represents the fraction of glutamate side chains involved in a multivalent close contact with two positively charged residues, which is 6-fold greater for arginine than for lysine (see Methods). (**I**) Mean-square displacement (MSD) of individual ProTα residues at increasing lag times shows that the lower friction (higher mobility) of uncharged residues resulting from weaker contacts is evident at short times, but is subsequently smoothed out at longer times when differences in friction for individual residues are averaged over longer chain segments. (**J**) Examples of exchange between lysine salt bridges in H1–ProTα (top) and arginine salt bridges in protamine–ProTα dense phases (bottom). Multivalent contacts* (62) *between negatively charged residues and arginine are more frequent and more stable than with lysine, as illustrated by representative snapshots from the simulations (see also Videos S1 and S2). (**K**) Lysine–glutamate contacts in the H1–ProTα dense phase exchange more frequently than arginine–glutamate contacts in the protamine–ProTα dense phase. Among the contacts that persist for over 100 ns, the majority do not remain intact continuously but break and reform between the same two residues (see Methods for details).*

The previous in-depth comparison of the simulations for ProTα and H1 with experimental observables, including protein concentrations, translational diffusion coefficients, intrachain distances, and chain dynamics, provided a validation of simulations with the force field and slab configuration employed (41). Moreover, the simulations at lower salt concentration and with protamine instead of H1 reproduce the higher protein concentrations (Figure S4) and the slower chain dynamics observed experimentally in droplets (Figure 3C, Figure S5), indicating that the force field also captures the effect of salt and differences in amino acid-specific interactions (30).

On average, each ProTα molecule in the dense phase is simultaneously in contact with ~6-7 H1 or ~11 protamine molecules, respectively (Figure 3D). Detailed information on the distribution of interactions between positively and negatively charged side chains in the resulting percolated network (63) can be obtained from contact profiles (Figure 3E) and contact maps (Figure S6). The average number of contacts that each residue in ProTα makes with other chains reveals remarkably similar interaction patterns in the dense phases with the different interaction partners, with local maxima at clusters of negatively charged residues in ProTα (41, 64). The absolute numbers of contacts, however, differ substantially between the different dense phases: The average number of contacts that ProTα residues make with protamine is ~80% greater than with H1, and ~10% greater at 8 mM than at 128 mM salt. The origin of the pronounced difference in interaction strength between lysine- and arginine-rich sequences in the simulations is expected to lie in the characteristic multipole of arginine (65), its weak hydration (66), and large polarizability (52), although especially the latter can only be captured indirectly with non-polarizable all-atom force fields (67).

The stronger interchain interactions at low salt and for arginine-rich sequences are thus likely to be at the root of the slower dynamics observed experimentally (Figure 2). Indeed, the average lifetime of contacts in the dense-phase simulations of protamine-ProTα is about 10 times longer than for H1-ProTα (Figure 3F). The duration of the contacts is in turn expected to be a determining factor for the motion of the polypeptide chain as a whole (29, 31, 68, 69). This expectation is corroborated by the correlation between contact lifetimes and chain reconfiguration times estimated from the simulations (Figure 3G) (see Methods for details). The similarity between simulated and measured reconfiguration times (Figure 3C) further suggests that the atomistic picture emerging from the MD simulations can help to explain the dynamics observed experimentally. Although complete equilibration of the protamine-ProTα simulations is challenging even with high-performance computing, the analysis is robust with respect to the quantities we report (Figure S4).

The simulations yield a picture in which charged residues form close contacts, as reflected by a pronounced short-range peak in the residue-residue distance distribution that is absent for uncharged residues (Figure 3H). This interaction is also reflected in the diffusion profile of charged residues, which at short times show a lower mobility than their uncharged neighbors (Figure 3I). However, these differences average out at longer times when the motion is dominated by larger chain segments. It is worth emphasizing that the contact lifetimes between individual charged residues are roughly two orders of magnitude shorter than the global reconfiguration times of the polypeptide chains. An important contribution to the short contact lifetimes is the rapid exchange between interacting side chains at the exceedingly high — roughly 1 M — concentrations of charged residues in the dense phases (41) (Figure 3J, K). Owing to the separation of timescales between contact lifetimes and the reconfiguration dynamics of entire chains, tens of thousands of residue-residue contacts are made and broken during $\tau_r$ (Figure 3F). Correspondingly, the different strengths of side chain interactions made by lysine and arginine can also be considered to result in different average frictional forces acting on the chains rich in lysine and arginine, respectively (70).

**Universal link between nanoscale, microscale, and mesoscale dynamics**

The effects of amino acid composition and salt concentration observed in the simulations, especially the correlation between contact lifetimes and reconfiguration times (Figure 3G), imply a quantitative link between side chain interactions and larger-scale motion, as previously suggested based on coarse-grained simulations (29, 31, 32, 68, 69). Striking linear correlations are also observed between the experimental chain reconfiguration times, translational diffusion times, and droplet viscosities (Figure 2J). Given these correlations across length- and timescales, we thus seek a physical model for condensate dynamics that allows us to predict microscale and mesoscale properties from nanoscale dynamics and vice versa.

Polymer physics presents an opportunity to conceptualize the dynamics of biomolecules in condensates across scales (Figure 4A-C). The residue–residue interactions observed in our simulations could be taken to suggest a model that explicitly accounts for the role of individual stickers mediating discrete binding events between the chains (71–73). However, existing quantitative models, such as the sticky Rouse model (72), typically assume bond energies greater than 10 $k_BT$* and correspondingly long contact lifetimes (74), and they have been developed primarily for systems with only a few stickers per chain, where chain dynamics between stickers occur faster than the contact lifetimes. These assumptions do not hold in our complex coacervates, where the nanosecond contact lifetimes suggest much lower effective bond energies, and where charged residues acting as stickers constitute a large fraction of the polymer — up to 100%. If the number of stickers approaches the total number of monomers in a chain, the sticky Rouse model converges to a standard Rouse model with increased monomer friction (74). Indeed, we are in this limit, with contact lifetimes orders of magnitude shorter than the chain reconfiguration times, so that thousands of contacts are made and broken along the chain during $\tau_r$ (Figure 3F). The effect of side chain interactions can then be captured by an effective friction coefficient, rather than explicitly accounting for bond formation dynamics.

This idea is used in the Rouse model of polymer solutions (24, 25, 75) (Figure 4B), which describes the dynamics of chains in terms of *N* connected segments subject to Brownian motion with a friction coefficient, $\zeta$. The resulting relation between the translational diffusion coefficient of the entire chain, $D = \frac{k_BT}{N\zeta}$, and the Rouse time of the chain, $\tau_R$ (roughly the reconfiguration time, $\tau_r$, see Methods), is

$$\tau_R = \frac{\langle R^2 \rangle}{3\pi^2 D}, \qquad \text{Eq. 1}$$

where $\langle R^2 \rangle$ is the mean squared end-to-end distance of the chain (see Methods). The bulk droplet viscosity, $\eta$, can be expressed in terms of the friction coefficient acting on the individual chain segments, and thus in terms of the experimental observables *D* and $\tau_R$ according to

$$\eta(D) = \eta_s + \frac{k_BT\, c_p \langle R^2 \rangle}{36} \frac{1}{D} \quad \text{and} \qquad \text{Eq. 2}$$

$$\eta(\tau_R) = \eta_s + \frac{\pi^2 k_BT\, c_p}{12} \tau_R, \qquad \text{Eq. 3}$$

where $\eta_s$ is the solvent viscosity, and $c_p$ is the protein concentration in the condensates (see Methods). Using the experimentally measured values of $\eta$, $c_p$, and $\tau_R$, the model correctly predicts — without any adjustable parameters — the linear dependencies observed experimentally, with absolute values within an order of magnitude of the experimental findings (Figure 4D, dashed lines). The model thus explains much of the mesoscopic properties of the droplets based on the notion that a polymer chain within the droplet behaves essentially like an isolated polymer within a more viscous medium imparting friction on the chain segments. The MD simulations support this notion based on the separation of timescales between contact lifetimes and chain reconfiguration and the large number of contacts made and broken during the reconfiguration time. The proportionality between contact lifetimes and chain reconfiguration times (Figure 3G) is additional evidence that friction depends on the duration of individual contacts.

However, based on the measured chain dimensions and protein concentrations, with average protein volume fractions between 17% and 31% (Figure 1E and Methods), the dense phase is in the semidilute regime (see Methods), where the chains partially overlap, indicating that interactions beyond purely frictional contributions may need to be taken into account. Indeed, the entanglement concentration is estimated to be in the range of the protein concentrations we observe in the dense phases (see Methods), suggesting that we are in a regime corresponding to the onset of entanglement effects between chains (76). This conclusion is further supported by the ~4$^{th}$-power dependence of viscosity on protein concentration (Figure S7A), and by the MD simulations, which indicate that every protein chain interacts with 6 to 11 other chains (Figure 3D), suggesting a contribution of entanglement-like effects that restrict the reorientation of the chains within the network of other chains (75, 77). Under these conditions, the experimentally observable chain reconfiguration time, $\tau_\mathrm{r}$, roughly corresponds to the disengagement time, $\tau_\mathrm{d}$ (Figure S7B, see Methods). Including entanglement in the Rouse model (Eq. 2 and 3) yields a correction to the expressions for viscosity by the factor $\frac{\langle R^2\rangle}{a^2}$ (75, 78), i.e.

---

* $k_B$, Boltzmann constant; *T*, temperature

$$\eta(D) = \eta_s + \frac{k_B T\ c_p \langle R^2\rangle}{36} \frac{\langle R^2\rangle}{a^2} \frac{1}{D} \text{ and} \qquad \text{Eq. 4}$$

$$\eta(\tau_\mathrm{d}) = \eta_s + \frac{\pi^2 k_B T\ c_p}{12} \frac{\langle R^2\rangle}{a^2} \tau_\mathrm{d}\ , \qquad \text{Eq. 5}$$

where $a$ is the diameter of a tube-like region made of the surrounding polymers within which the motion of the chain is essentially confined (75, 79) (Figure 4C). This effective tube diameter (79) is expected to be on the order of the mesh size (75), which we previously estimated to be ~2–4 nm (41), as well as the correlation length (80), which is ~1–5 nm (see Methods). An analytical estimate based on protein concentration and chain dimensions (81) yields $a$ = 3 ± 1 nm (see Methods). If we treat $a$ as an adjustable parameter, we obtain a value of 4 ± 2 nm from a fit of Eq. 4 to the experimental data (Figures 4D, S7C), in remarkable agreement with these estimates. The value of $a$ is only about half of $\langle R^2\rangle^{1/2}$, confirming that the polypeptides are only weakly entangled, as expected for such short chains (75). Nevertheless, the contribution of entanglement is essential for quantitative agreement with experiment (Figure 4D). It is worth emphasizing that Eqs. 1–5 are relations between quantities that are measured independently, and that the predictions have no adjustable parameters except $a$, which turns out to match independent estimates for our coacervates (81).

The agreement with the Rouse model across all our coacervates prompts the question of whether its applicability is limited to complex coacervates of highly charged proteins. We thus compared with the behavior of four other phase-separated systems for which at least some of the pertinent data are available: three proteins that form biomolecular condensates and for which diffusion coefficients and bulk viscosities have been reported, LAF-1 (59), A1-LCD (33, 82),[†] and FUS-LCD (83, 84); and a synthetic complex coacervate widely used in industrial applications, consisting of PMAPTAC (poly([3-(methacrylamido)propyl]trimethylammonium chloride)) and PAMPS (poly(2-acrylamido-2-methyl-1-propanesulfonic acid) (85), with bulk viscosities and disengagement times[‡] roughly six orders of magnitude greater than those of the biomolecular systems. Remarkably, all of these data are in line with the behavior of the coacervates studied here (Figure 4D), suggesting that the Rouse framework is more generally applicable and may provide a simple universal link between nanoscopic, microscopic, and mesoscopic properties of biomolecular condensates formed by disordered proteins and even synthetic polyelectrolytes. As a result, we can also provide order-of-magnitude estimates for the expected end-to-end reconfiguration times of the other IDPs in their dense phases: approximately 0.5–5 μs for A1-LCD, 3–30 μs for FUS-LCD, and 10–100 μs for LAF-1.

[†] The diffusion coefficient was estimated from the radius of gyration, the diffusion time through the confocal volume, and the size of the confocal volume.

[‡] The disengagement time, roughly corresponding to the chain reconfiguration time (see Methods), or reptation time (25), was inferred from the crossover frequency in rheological measurements. The agreement of all data with Eq. 5 (inset in Figure 4D) highlights the consistency between FRET-based nsFCS for direct measurements of chain dynamics and the indirect determination of those dynamics from the viscoelastic properties of polymer solutions.

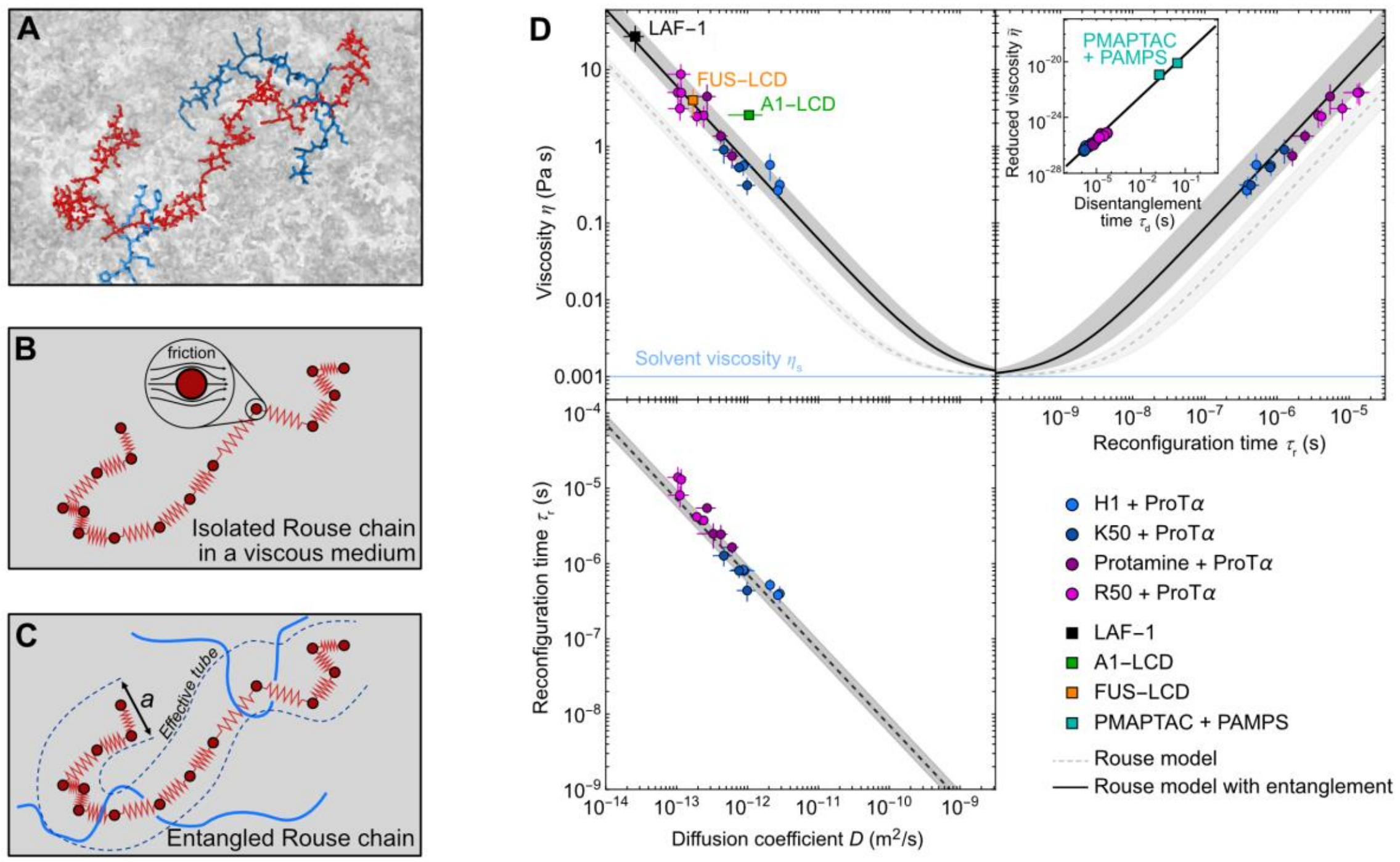

***Figure 4. Polymer models provide a quantitative link between single-chain dynamics, translational diffusion, and bulk viscosity.*** *(**A**) Illustration of a ProTα molecule (red) in the H1-ProTα dense phase (gray) from MD simulations, with two H1 segments entangled with ProTα shown in blue. (**B**) Schematic of the Rouse model corresponding to **A**, with beads (red circles) subject to Brownian motion and friction from the environment, and entropic springs connecting them. (**C**) Schematic of the Rouse model with entanglement, where the motion of a Rouse chain is constrained by a network of obstacles with a characteristic distance between them equal to $a$, the effective tube diameter* (79) *or entanglement spacing* (78)*. (**D**) Comparison of the experimentally observed viscosities, diffusion coefficients, and chain reconfiguration times with the prediction of the Rouse model (dashed line) and the Rouse model with entanglement (solid line) from Equations 1-5 and $a = 4 \pm 2\,nm$ (see Methods), including the viscosities and diffusion coefficients of LAF-1* (59)*, A1-LCD* (33, 82)*, and FUS-LCD* (83, 84)*. Note that the value of $a$ for these condensates may be different from the coacervates investigated here, therefore the model predictions are only indicative for these systems. The inset reports the relation in Eq. 5 between the disengagement time, $\tau_\mathrm{d}$ (roughly corresponding to $\tau_\mathrm{r}$, see Methods), and the reduced viscosity, $\tilde{\eta} = (\eta(\tau_\mathrm{d}) - \eta_s)a^2 / \, c_p\langle R^2\rangle$ in Pa s m³, which allows us to compare our samples with PMAPTA + PAMPS (chain lengths P619 and P1188)* (85)*. The error bands of the fits account for the differences in $a$ (Figure S7C), $\langle R^2\rangle$, and $c_p$ between the different samples. Data are presented as mean values ± s.d. (see Methods).*

## Discussion

Our results demonstrate a close mechanistic link between interaction dynamics across scales: from the contact lifetimes between amino acid residues and the resulting chain dynamics at the molecular scale, to the micro- and mesoscale dynamics and viscosity of biomolecular condensates formed by charged disordered proteins. Both the amino acid composition and the salt concentration modulate the interactions and dynamics within these complex coacervates, with a particularly pronounced role for the charge density in the chains and their arginine content, which has previously been found to be an important residue for driving biomolecular phase separation (50, 62). Key quantities reflecting the dynamics of condensates, ranging from the chain reconfiguration time and the translational diffusion of chains to the bulk viscosity of the condensates, can be linked quantitatively by the Rouse model with entanglement. Existing data on condensates whose formation is driven by hydrophobic interactions as well as complex coacervates formed by synthetic polymers suggest the existence of the same mechanistic link between scales in those systems. The success of the Rouse model for such a wide range of different biomolecules indicates that key aspects of the underlying physics of these systems are remarkably similar.

As expected from the high protein concentrations inside the condensates, and as indicated by the MD simulations of the complex coacervates investigated here as well as by previous simulations (31, 63, 86, 87), the protein chains form a highly connected network of interactions, the hallmark of viscoelastic network fluids. Despite the expected viscoelasticity of such systems, we observe the viscous component of the shear relaxation modulus to be dominant for the complex coacervates formed from highly charged disordered proteins on the accessible timescales, both in the present work and in previous results on H1 and ProTα (41). For instance, the microrheological measurements by bead tracking are well described in terms of normal Brownian diffusion down to the shortest accessible timescales in the millisecond range (Figure S3B-D); for H1-ProTα, droplet relaxation upon fusion is single-exponential, with a relaxation time proportional to the radius of the final droplet, which also indicates that the viscoelasticity of the dense phase on the millisecond timescale and above is dominated by a viscous (rather than an elastic) component (20, 21). The MD simulations of H1-ProTα (41) and protamine-ProTα condensates (Figure 3) suggest an interesting molecular mechanism contributing to the pronounced fluidity of these complex coacervates: The extreme concentration of charged side chains of >1 M in the dense phase, corresponding to an average distance between charged groups of <1 nm, facilitates the formation of transient ternary interactions between multiple charged groups (Figure 3H). These interactions lead to the rapid exchange of contacts between residues (Figure 3J) (41). This type of dynamic shuffling may be essential for many processes in the cell, e.g., to prevent dynamic arrest in compartments such as the cell's nucleus, which is densely packed with highly charged polyelectrolytes (36, 88).

The abundant evidence for an elastic contribution to stress relaxation[§] in other biomolecular condensates (26, 60, 89, 90) raises the question of why viscous relaxation dominates for the coacervates we studied. To address this point, we estimated the frequency dependence of the loss and storage moduli according to Rouse theory based on our experimentally defined parameters (24, 75, 80). The crossover frequency of the two moduli is predicted to be in the range of the inverse chain reconfiguration time (Figure S8). For the coacervates investigated here, the elastic component is thus expected to be dominant only on timescales in the microsecond range and below, and would thus require megahertz microrheology to be detected (91). Our observations and the relations we propose are even in agreement with measurements on synthetic complex coacervates whose reconfiguration times of ~1 s[‡] correspond to bulk viscosities six orders of magnitude above those of our protein-based systems (85). Therefore, biomolecular condensates with pronounced elastic relaxation at lower frequencies (26, 60, 89, 90) would be expected to show correspondingly slower chain reconfiguration and much higher viscosities.

We note, however, that there are several types of biomolecular condensates that are not expected to be described by simple polymer physics, for instance condensates composed predominantly of structured molecules (92), such as folded proteins or RNA; condensates driven by interactions between folded domains mediating long-lasting crosslinks between molecules (93), whose viscoelastic moduli are thus expected to be dominated by the timescales for making and breaking crosslinks (94); or condensates that form persistent intermolecular structures or exhibit aging (33, 60). Especially the latter systems can form kinetically arrested aggregates and rigid solids (95–97), whose persistent structure and nonequilibrium properties will require residue-specific interactions and desolvation effects to be accounted for (33, 98–100). Likewise, simulations of protein–nucleic acid condensation and of systems with high charge patterning suggest the formation of local clustering (101) and caged dynamics (102), behaviors not observed in the homogeneous condensates we investigate here. In those cases, the viscoelastic behavior is expected to be more complex than described by simple polymer models.

---

[§] The relative magnitude of viscous and elastic contributions to a material's mechanical response is timescale-dependent, with viscous behavior typically dominating at low frequencies, and elastic behavior at high frequencies. However, numerous reports describe biomolecular condensates exhibiting a dominant elastic contribution even at low frequencies relative to the timescales of dynamics at the molecular scale.

Two important factors contribute to the success of the simple mean-field Rouse framework for predominantly viscous and isotropic condensates: One is the pronounced separation of timescales between contact lifetimes and overall chain dynamics; the resulting time averaging over vast numbers of contacts makes the concept of friction applicable on the timescale of chain reconfiguration. Another is the absence of pronounced sequence patterns in the proteins and polypeptides included in our analysis; as a result of the corresponding effective spatial averaging (Figure 3I), a homopolymer model seems to provide a reasonable approximation. However, we do observe how some sequence-specific aspects govern molecular dynamics and material properties: condensates rich in arginine exhibit dynamics roughly two orders of magnitude slower than those rich in lysine[**]. Nevertheless, the agreement between theory and experiments across the systems suggests that the relations linking material properties and molecular dynamics serve as good first-order approximations. While these relations allow for the prediction of molecular dynamics from viscosity—and vice versa—they do not, on their own, enable the inference of either property solely from sequence composition.

Computational approaches offer the most promising path towards addressing this limitation, for instance by quantitatively relating the energetics and dynamics of molecular simulations to the viscoelasticity of condensates (31). An et al. (32) have reported that increased condensate stability correlates with low mobilities and high viscosities in coarse-grained simulations and employed active learning to identify the influence of amino acid composition and sequence patterning on the dynamic and thermodynamic properties of biomolecular condensates. We also observe correlations between some thermodynamic and dynamic properties for the condensates investigated here (Figure S2B-D). It has further been suggested that the nature of the contacts formed at the residue level can be related to viscoelastic properties via the eigenvalue spectra of Rouse-Zimm matrices that account for intra- and intermolecular contacts in the Rouse model, albeit not yet in terms of absolute timescales (33, 103–105). Using experimentally validated atomistic explicit-solvent simulations, as presented here, may enable the development of predictive approaches for absolute molecular timescales, which, according to our results, can be linked quantitatively to material properties at the mesoscopic scale.

**Methods**

ProTα was recombinantly expressed, purified, and labeled for single-molecule FRET experiments as described previously (41); human histone H1.0 was purchased from New England Biolabs (product code M2501S). Poly L-lysine hydrochloride (referred to as K50) and poly L-arginine hydrochloride (referred to as R50) were from Alamanda Polymers (Huntsville, AL, USA; catalog numbers 000-KC050, 000-R050). Protamine was from Sigma-Aldrich (product number P4020). The details of single-molecule experiments, concentration, viscosity, diffusion coefficient measurements, and the experimental setup have been described before (41, 64). See Supporting Information for detailed descriptions on protein expression, purification, labeling, experimental procedures, and analysis, as well as a description of the theory and simulations.

[**] In contrast, our results indicate no significant difference between the contact dynamics of glutamate and aspartate (Figure S6).

**Acknowledgments**
We thank Mark Nüesch for help with data analysis, Andreas Vitalis for help with CAMPARI, and Priya Banerjee, Sam Cohen, Kingshuk Ghosh, Hagen Hofmann, William Jacobs, Dmitrii Makarov, Murugappan Muthukumar, Rohit Pappu, Michael Rubinstein, Joseph Schlenoff, Andrea Soranno, and Valerio Sorichetti for insightful discussion. This work was supported by the Swiss National Science Foundation (310030_197776, B.S.), the Novo Nordisk Foundation Challenge program REPIN (#NNF18OC0033926, B.S.), the Intramural Research Program of the National Institute of Diabetes and Digestive and Kidney Diseases at the National Institutes of Health (R.B.B.), the Forschungskredit of the University of Zurich (N.G. and M.T.I.), the Ernst Hadorn Foundation (N.G.), and the European Union's Horizon 2020 research and innovation programme under the Marie Skłodowska-Curie grant agreement ID 898228 (A.C.). We utilized the computational resources of Piz Daint, Alps and Eiger at the CSCS Swiss National Supercomputing Centre, and of the National Institutes of Health HPC Biowulf cluster (http://hpc.nih.gov). Mass spectrometry was performed at the Functional Genomics Center Zurich. FRAP and bead tracking were performed with support of the Center for Microscopy and Image Analysis, University of Zurich.

## Supporting Information for

# Material properties of biomolecular condensates emerge from nanoscale dynamics

Nicola Galvanetto[1,2,*], Miloš T. Ivanović[1,*] , Simone A. Del Grosso[1], Aritra Chowdhury[1], Andrea Sottini[1], Daniel Nettels[1], Robert B. Best[3,*] and Benjamin Schuler[1,2,*]

[1]Department of Biochemistry, University of Zurich, Zurich, Switzerland
[2]Department of Physics, University of Zurich, Zurich, Switzerland
[3]Laboratory of Chemical Physics, National Institute of Diabetes and Digestive and Kidney Diseases, National Institutes of Health, Bethesda, MD, USA
* Corresponding authors: N. Galvanetto (n.galvanetto@bioc.uzh.ch), M. T. Ivanović (m.ivanovic@bioc.uzh.ch), R. B. Best (robert.best2@nih.gov), B. Schuler (schuler@bioc.uzh.ch)

**This PDF file includes:**

**Other supporting materials for this manuscript include the following:**

Video S1 to S2

## Supporting Information Text

## Materials and Methods

*Sample preparation and labeling*

We used recombinant human histone H1.0 (H1; New England Biolabs, product code M2501S). Poly L-lysine hydrochloride (referred to as K50, MW = 8200 Da, degree of polymerization 45–55) and poly L-arginine hydrochloride (referred to as R50, MW = 9600 Da, degree of polymerization 45–55) were from Alamanda Polymers (Huntsville, AL, USA; catalog numbers 000-KC050, 000-R050). Protamine was from Sigma-Aldrich (product number P4020). Labeled and non-labeled ProTα were prepared as described previously (106). Labeling was achieved by introducing cysteine residues at positions 56 and 110 for attaching fluorophores (detailed protein sequences in Extended Data Table 1). Prior to labeling, the protein was reduced in phosphate-buffered saline (PBS), pH 7, containing 4 M guanidinium hydrochloride (GdmHCl) and 0.2 mM EDTA, using 10 mM Tris(2-carboxyethyl) phosphine hydrochloride (TCEP) for 60 minutes, followed by multiple (5x) buffer exchange steps to a similar PBS solution without TCEP (pH 7, 4 M GdmCl and 0.2 mM EDTA) using centrifugal filters with a 3-kDa molecular mass cutoff (Sigma-Aldrich). Labeling was achieved with Cy3B maleimide (Cytiva) and CF660R maleimide (Sigma-Aldrich) at a protein: dye ratio of 1:6:6, incubated for an hour at room temperature and overnight at 277 K. Excess dye was reacted using 10 mM dithiothreitol for ten minutes and removed by centrifugal filtration. The labeled protein was subsequently purified by reversed-phase high pressure liquid chromatography using a Reprosil Gold C18 column (Dr. Maisch, Germany), without separating the labeling permutants. The correct mass of labeled protein was confirmed by electrospray ionization mass spectrometry.

*Turbidity measurements*

To assess the extent of phase separation, the relative turbidity was quantified by the attenuation of light at 350 nm with a NanoDrop 2000 UV-Vis spectrophotometer (Thermo Scientific). The positively charged polypeptides were added to a fixed volume of a ProTα solution to achieve a final concentration of 10 µM ProTα and investigate a wide range of stoichiometric ratios. These experiments were carried out in TEK buffer, composed of 10 mM Tris-HCl, 0.1 mM EDTA (pH 7.4), with ionic strength adjusted using KCl. Samples were rapidly mixed via pipetting for approximately 10 seconds before measurements. Each sample was measured four times, and the results were averaged. Both protein stock solutions were diluted in identical buffers before the mixing process.

*Single-molecule fluorescence spectroscopy*

We performed confocal single-molecule analysis, concentration determination, and fluorescence correlation spectroscopy at 295 K with a MicroTime 200 (PicoQuant), equipped with a UPlanApo 60×/1.20-W objective (Olympus), mounted on a piezo stage (P-733.2 and PIFOC from Physik Instrumente GmbH), using a 532-nm continuous-wave laser (LaserBoxx LBX-532-50-COL-PP; Oxxius), and a 635-nm diode laser (LDH-D-C-635M; PicoQuant). Fluorescence photons, which were separated from scattered laser light using a triple-band mirror (zt405/530/630rpc from Chroma), were initially separated into two channels by either a polarizing or a 50/50 beam splitter, and then into four detection channels with dichroic mirrors to separate donor and acceptor emission (T635LPXR from Chroma). Donor emission was further filtered with an ET585/65m band-pass (Chroma), and acceptor emission with an LP647RU long-pass filter (Chroma), before being detected by SPCM-AQRH-14-TR single-photon avalanche diodes (PerkinElmer). SymPhoTime 64 version 2.4 (PicoQuant) was used for data acquisition.

In single-molecule measurements, ProTα, labeled with Cy3B and CF660R, was excited by the 532-nm laser. Experiments were conducted in TEK buffer, including different concentrations of KCl. Plastic sample chambers (µ-Slide, ibidi) were used to mitigate surface adhesion of the positively charged polypeptides to glass. For measurements of dilute-phase samples, the power of the 532-nm laser was set to 100 µW (measured at the back aperture of the microscope objective); the confocal volume was positioned 30 µm deep into the sample chamber; and concentrations of labeled protein between 50 and 100 pM were used. For single-molecule measurements in the dense phase, the average power at the back aperture was between 3 and 30 μW for continuous-wave excitation, depending on the background level; the confocal volume was placed at the center of the spherical droplets, whose radii varied between 4 and 30 μm. Unlabeled proteins (12 μM ProTα and a concentration of the positively charged polypeptides to ensure charge balance) were mixed with 5 to 10 pM of labeled ProTα. Photon bursts, occurring as proteins traverse the confocal volume, were isolated from background-corrected fluorescence trajectories, binned at 4 ms, with a photon count threshold of 285 per burst. In dilute-phase measurements, bursts were defined as sequences of at least 285 consecutive photons with interphoton times below 40 μs.

Ratiometric transfer efficiencies were obtained from $E = N_A/(N_A + N_D)$, where $N_A$ and $N_D$ are the numbers of donor and acceptor photons, respectively, in each photon burst, corrected for background, channel crosstalk, acceptor direct excitation, differences in quantum yields of the dyes, and detection efficiencies (107, 108). Mean transfer efficiencies, $\langle E \rangle$, were determined from fits with Gaussian peak functions to the transfer efficiency histograms. To infer dye-to-dye distance distributions, *P*(*r*), from $\langle E \rangle$, we use the relation (58)

$$\langle E \rangle = \langle \varepsilon \rangle \equiv \int_0^\infty \varepsilon(r) P(r) dr, \qquad \text{Eq. 6}$$

where

$$\varepsilon(r) = R_0^6/(R_0^6 + r^6). \qquad \text{Eq. 7}$$

The Förster radius, $R_0$, (109) of 6.0 nm for Cy3B/CF660R in water (110) was corrected for the refractive index, *n*, in the droplets according to the published dependence of *n* on the protein concentration (111), which is linear up to a mass fraction of at least 50 % (112) and only marginally dependent on the type of protein (111). At the dense-phase protein concentrations observed here, *n* is greater than the value in water by 3%–6%, resulting in a slightly smaller $R_0$ inside the droplets (5.8–5.9 nm). Systematic uncertainties in transfer efficiencies due to instrument calibration and uncertainty in $R_0$ are estimated to be roughly ±0.03, in line with the range previously reported (107). The precision for measurements on the same instrument is higher, typically <0.01 (110). *P*(*r*) was estimated using the length scaling exponent $\nu$ by applying an empirically modified self-avoiding-walk polymer (SAW-$\nu$) model (113, 114). To estimate the mean square end-to-end distance of the complete ProTα chain, we used $\nu$ and the total number of amino acids, $N_{tot}$ = 110. The impact of fluorophore labeling on ProTα-H1 interaction was minimal, as evidenced by previous studies (64, 106). Given the fraction of <$10^{-6}$ of labeled protein in the dense phases, effects of fluorophore labeling on dense-phase behavior were considered negligible. Analysis of fluorescence data was performed with the software package Fretica (https://github.com/SchulerLab) run with Mathematica 12.3 (Wolfram Research).

*Measurements of protein concentrations and diffusion coefficients in the dilute and dense phases*
We utilized fluorescence correlation spectroscopy (FCS) and quantitative fluorescence intensity analyses using a MicroTime 200 (PicoQuant) to assess the concentrations of ProTα molecules doubly labeled with Cy3B and CF660R, in the dense and dilute phases (41, 82). A mixture of unlabeled proteins (12 μM ProTα and a concentration of the respective positively charged polypeptides to ensure charge balance), doped with a small concentration (~10 pM to 10 nM) of labeled ProTα in TEK buffer including the specified concentrations of KCl was allowed to phase-separate at 295 K. To analyze the dilute phase, the phase-separated mixture was centrifuged at 295 K for 30 minutes at 25,000 g, leading to the formation of a single large droplet of dense phase. The supernatant was then aspirated and placed into a sample chamber (μ-Slide, ibidi). For measurements in the dense phase, the phase-separated mixture was directly transferred to the sample chamber, and droplets were allowed to settle on the bottom surface of the sample chamber by gravity; the boundaries of individual droplets were identified via 3D confocal imaging, and FCS and intensity measurements were performed by focusing inside the droplets.

To excite CF660R, we employed the 635-nm continuous-wave laser at 5 μW (measured at the back aperture of the objective). The emitted fluorescence photons were then separated using a polarizing beam splitter and subsequently detected by two detectors. The collected correlation data were analyzed employing a model that assumes a 3D Gaussian-shaped confocal volume:

$$G(\tau) = G_0\left[\left(1+\frac{\tau}{\tau_D}\right)\sqrt{1+s^{-2}\frac{\tau}{\tau_D}}\right]^{-1}, \qquad \text{Eq. 8}$$

where $\tau$ is the lag time, $G_0$ is the amplitude, $\tau_D$ is the translational diffusion time, and $s$ is the ratio of the axial to lateral radii of the confocal volume. The calibration curve was generated from the analysis of samples with known concentrations (0.3, 1, 3 10, 30, and 100 nM) of labeled ProTα in TEK buffer including 120 mM KCl.

Concentrations were estimated from the average number of labeled proteins in the confocal volume, $N_p = \left(1-\frac{n_b}{n_f}\right)^2/G_0$, as previously described (82), where $n_b$ is the background count rate estimated from samples without labeled protein, and $n_f$ is the average count rate of the measurement with labeled ProTα. As an alternative method for estimating concentrations, the fluorescence intensity after background subtraction was used based on a corresponding calibration curve. Total ProTα concentrations were obtained by dividing the concentrations of labeled ProTα by the doping ratio, which was set to ensure that fluorescence intensities fell within the range of linear response of detection. This approach requires higher doping ratios for measurements in the dilute phase than for those in the dense phase. For each set of experimental conditions, a minimum of two concentration estimates were made, one using FCS and one using intensity detection. These assessments were repeated a minimum of two times to increase reliability.

Diffusion coefficients were calculated from translational diffusion times, $\tau_D$, using a calibration curve

$$D = \frac{\Lambda}{\tau_D}, \qquad \text{Eq. 9}$$

where $\Lambda$ was obtained from a calibration with samples of known diffusion coefficient in water. The calibration was cross-validated in ProTα-H1 droplets with two-focus FCS (115) to account for effects of refractive index differences between dilute and dense phase on the observed translational diffusion coefficients (41).

Since maximum dense phase formation occurs at a mixing ratio close to charge balance (Figure S1), all experiments were performed by mixing ProTα and the positively charged polypeptides at this ratio. Since reproducible droplet formation becomes difficult to maintain and exceedingly sample consuming at salt concentrations near the critical point, a compromise between experimental feasibility and accessible salt concentrations was made.

*Microrheology*

We mixed ProTα and the positively charged polypeptides, both unlabeled, under phase separating conditions at charge balance with an aliquot of fluorescent microspheres (100 nm and 500 nm diameter, Fluoro-Max, Thermo Fisher Scientific). After centrifugation, we collected a single large droplet (diameter ≥100 µm) for observation. The motion of the beads within the droplet was tracked at 295 K using an Olympus IXplore SpinSR10 microscope equipped with a 100×/1.46 NA plan-apochromat oil immersion objective, for 300 seconds with 50 ms exposure time per image and and acquisition rate of 5 Hz. Bead trajectories were obtained using the TrackMate plugin of Image J (116) and further analyzed with MATLAB 2016b (MathWorks). We calculated the mean square displacements (MSD) in the image plane, averaging across 20 trajectories, to obtain the diffusion coefficient, *D*, from ⟨*MSD*(*t*)⟩ = 4*Dt*, where *t* is the time after the start of observation.

The MSD analysis demonstrated uniform viscous properties within the droplets, as evidenced by the Brownian diffusion, the consistency between different beads probed in the droplet volume, and the uniform fluorescence intensity observed in microscopy images. We did not observe aging effects on the timescale up to days. We estimated the viscosity, $\eta$, using the Stokes–Einstein equation assuming freely diffusing Brownian motion of particles with hydrodynamic radius $R_h$:

$$\eta = \frac{k_B T}{6\pi D R_h}. \quad \text{Eq. 10}$$

In this study, we were interested in the bulk viscosity of the medium that can be measured by probes that are much larger than the correlation length of the polymer network (117, 118). At short times, some MSD curves apparently deviate from the linear Brownian behavior. This effect is due to uncertainties in position determination owing to out-of-focus beads (Figure S3C, D), rather than due to possible elastic properties of the dense phase, which cannot be resolved within the time scales studied here (80) (Figure S8).

To increase the time resolution of tracking, we also tested K50-ProTα droplets in an optical tweezers instrument (C-Trap, LUMICKS, Amsterdam). K50-ProTα was the sample least prone to photodamage by the IR laser. A condensate-forming sample (3 µl) mixed with 1-µm polystyrene beads was placed on a polymer coverslip (ibidi GmbH, Germany) at the center of an enclosure formed by double-sided tape. Another polymer coverslip was placed on top of the sample, sandwiching and sealing it. The condensate sample was left to equilibrate for 30 min and then placed on the sample stage of the optical tweezers instrument equipped with a 60× water immersion objective and a bright-field camera. We trapped isolated beads within large droplets (diameter > 50 µm) with minimal laser power to prevent photodamage and the formation of optically visible bubbles at the bead surface. Bead motion was recorded with the camera at acquisition rate >300 Hz and tracked with Blulake (LUMICKS, Amsterdam). MSDs were calculated in Mathematica (Wolfram Research).

*Correlation length, overlap concentration, entanglement concentration, and effective tube diameter*

The correlation length in the dense phase was estimated from $\xi \approx R_g \left(\frac{c_p}{c^*}\right)^{-\frac{\nu}{3\nu-1}}$ (25), where $c_p$ represents the total protein concentration (concentration of ProTα plus the concentration of the positively charged partner); $c^*$ is the overlap concentration, defined as $c^* = 1/V$ with $V$ approximating the volume occupied by a polymer chain ($V \approx 4/3\,\pi R_g^3$); $R_g$ is the radius of gyration ($R_g \approx \sqrt{\langle R^2 \rangle}/\sqrt{6}$, where $\sqrt{\langle R^2 \rangle}$ is the mean square end-to-end distance of the polymer); and $\nu$ is the scaling exponent. The condition $c_p \gtrsim c^*$ corresponds to the transition from the dilute to the semidilute regime, marking the beginning of chain interpenetration (25, 119). The concentrations corresponding to $c^*$ in our samples range from 7 mM to 10 mM, while the protein concentrations corresponding to $c_p$ in the four samples range from 14 mM to 42 mM (concentration averages at the different protein compositions and salt concentrations). The correlation lengths, $\xi$, are between 1 nm and 5 nm; this range was estimated using the average values of $R_g$ and $c_p$ for the four protein combinations at the different salt concentrations; values of $\nu$ ranging from 0.5 to 0.6; and values of $c^*$ based on the lower and upper bounds of both $R_g$ and the theoretical limits for the hydrodynamic radius ($R_g/1.5 < R_h < R_g/0.77$) (25, 120) as chain dimensions. Obtaining a precise measured value of $\nu$ is challenging; from internal length scaling of the disordered proteins in the MD simulations (120), we estimate values of $\nu$ ranging from 0.53 to 0.63 for the different coacervates and salt concentrations. Since $c_p > c^*$ for all conditions used, the correlation lengths are thus smaller than the chain dimensions, so we expect excluded volume interactions to be screened by the overlapping chains on the length scales we are probing experimentally (25). Similarly, hydrodynamic interactions are expected to be screened under these conditions, suggesting that Rouse dynamics apply on the length scales of entire chains (25, 119). For consistency and simplicity, we thus use the corresponding value of $\nu$ = ½ throughout.

The entanglement concentration, $c_e$, delineates the transition from a non-entangled polymer solution to an entangled one when $c_p \gtrsim c_e$. This condition is met approximately if the protein concentration is high enough that there are at least two chains in the volume $V$ pervaded by a single chain (see eq. 10 in ref. (76)), yielding $c_e \approx 2c^*$. As $c_p \approx c_e$ in our systems, only mild entanglement effects are expected. The effective tube diameter, $a$, was estimated from the experimental protein concentration, $c_p$, and chain dimensions, $\langle R^2 \rangle$, following ref. (81):

$a = l_k \sqrt{\alpha \left(\frac{l_k}{\rho_m {l_0}^3 p}\right)^{-2} + \beta \left(\frac{l_k}{\rho_m {l_0}^3 p}\right)^{-1} + \gamma \left(\frac{l_k}{\rho_m {l_0}^3 p}\right)^{-1/3}}$, where $l_0 = 0.4\,\mathrm{nm}$ is the backbone bond length, $l_K = \frac{\langle R^2 \rangle}{L_c} \approx 1.8\,\mathrm{nm}$ is the Kuhn length, $L_c$ is the contour length of ProTα, $p = (c_p \langle R^2 \rangle)^{-1}$ is the packing length, $\rho_m \approx c_p L_c / l_0$ is the monomer number density, and $\alpha$, $\beta$, and $\gamma$ are obtained from the table in ref. (81). Considering the different protein concentrations and chain sizes in our complex coacervates, and the two different sets of values for $\alpha$, $\beta$, and $\gamma$ provided (81), we obtain $a = 3 \pm 1\,\mathrm{nm}$.

Total volume fractions of protein, $\phi$, were estimated from the measured ProTα concentrations, $c_{\mathrm{ProT\alpha}}$ (Figure 1E), as $\phi = (M_{ProT\alpha}\, c_{ProT\alpha} \bar{v}_{ProT\alpha} + X\, M_{partner}\, c_{ProT\alpha}\, \bar{v}_{partner})$, where $M$ is the molar mass, $\bar{v}$ is the partial specific volume, and $X$ is the mixing ratio between the positively charged partner and ProTα to reach charge balance. We note that since the samples examined here are complex coacervates of similar but not identical lengths, the values must be considered approximate.

*Nanosecond fluorescence correlation spectroscopy (nsFCS)*

Samples for *ns*FCS measurements were prepared as outlined in the section *Single-molecule measurements*. To avoid the reduction in signal due to photobleaching of slowly diffusing molecules in the dense phase, we moved the confocal volume at constant speed (3 µm/s) following a serpentine trajectory in a horizontal plane within the droplet (41). Excitation with continuous-wave laser light of 532 nm was conducted at 3 or 30 µW (measured at the back aperture of the objective). For subpopulation-specific correlation analysis of the FRET-active species, we used photons from bursts with $\langle E \rangle > 0.15$. Autocorrelation curves of acceptor ($A$) and donor ($D$) detection channels, and cross-correlation between *A* and *D* were analyzed as described previously (41, 121). In Figure 2E we show cross correlation curves with logarithmically spaced lag times ranging from nanoseconds to milliseconds. The function used for fitting the correlations between detection channels $i, j = A, D$ was

$$G_{ij}(\tau) = G_{0,ij}\ \frac{(1 - c_{ab}^{ij}\, e^{-|\tau|/\tau_{ab}^{ij}})(1 + c_{cd}^{ij}\, e^{-|\tau|/\tau_{cd}})(1 + c_{rot}^{ij}\, e^{-|\tau|/\tau_{rot}})}{\left(1 + \frac{|\tau|}{\tau_D}\right)\left(1 + s^{-2}\frac{|\tau|}{\tau_D}\right)^{1/2}}. \qquad \text{Eq. 11}$$

The three terms in the numerator with amplitudes $c_{ab}^{ij}$, $c_{cd}^{ij}$, $c_{rot}^{ij}$, and correlation times $\tau_{ab}^{ij}$, $\tau_{\mathrm{cd}}$, $\tau_{\mathrm{rot}}$ describe photon antibunching, conformational dynamics, and dye rotation, respectively. $\tau_{\mathrm{D}}$ and *s* are defined as in Eq. 8. Conformational dynamics result in a characteristic pattern with a positive amplitude in the autocorrelations ($c_{cd}^{DD} > 0$ and $c_{cd}^{AA} > 0$) and a negative amplitude in the cross-correlation ($c_{cd}^{AD} < 0$), but with a common correlation time, $\tau_{cd}$. The three correlation curves $G_{DD}(\tau)$, $G_{AA}(\tau)$, and $G_{AD}(\tau)$ were fitted globally with $\tau_{\mathrm{cd}}$ and $\tau_{\mathrm{rot}}$ as shared fit parameters. The relaxation time $\tau_{\mathrm{cd}}$ was converted into the chain reconfiguration time $\tau_r = \int_0^\infty \frac{\langle r(0) r(\tau) \rangle}{\langle r^2 \rangle} d\tau$, according to the procedure previously established (122). This conversion is based on the assumption that dynamics of the inter-dye distance $r$ can be represented as diffusive motion in a potential of mean force obtained from the distance distribution $P(r)$ by Boltzmann inversion (57, 122).

The experimental uncertainties of $\tau_{\mathrm{r}}$ reported here are either standard deviations calculated from three measurements or the error of the fit of $\tau_{\mathrm{cd}}$ (from which $\tau_{\mathrm{r}}$ is derived), whichever was greater. We estimate the error of the fit from the variability resulting from fits using different lag-time intervals of the FCS data: We report as uncertainties the range of reconfiguration times obtained by using values from 0.8 ns to 8 ns as lower bounds, and from 1 ms to 6 ms as a upper bounds of the fitting window. The rotational correlation time, $\tau_{\mathrm{rot}}$, is approximately one order of magnitude smaller than the chain reconfiguration time, $\tau_{\mathrm{r}}$, and is caused by dye rotation (41).

*Polymer models*

We compared the experimental results of ProTα reconfiguration times, $\tau_r$, ProTα translational diffusion coefficients, $D$, in condensates, and bulk viscosities $\eta$ of the condensates, with corresponding values derived from three models commonly used for polymer solutions (Figure S7B): i) the Rouse model (most commonly used for concentrated solutions and melts) (24), ii) the Zimm model (most commonly used for dilute solutions) (123), and iii) the Rouse model with entanglement (75, 77, 79). Since the concentrations and chain dimensions in the dense phases are comparable in all systems studied, we tested the models using average values of these quantities rather than reduced quantities, which would not have allowed a comparison of different models with different dependencies on protein concentrations and chain dimensions for all samples at the same time. The upper and lower limits of the error bands in Figure 4D correspond to models calculated for the minimum and maximum values of the chain dimensions and protein concentrations in the set of experimentally observed values.

i) The Rouse model describes the dynamics of a chain of *N* beads connected by harmonic springs with root mean square length *b*, and subjected to Brownian motion with friction coefficient $\zeta$. The friction coefficient of the whole chain is approximated as $N\zeta$. The translational diffusion coefficient of the center of mass of the entire chain, *D*, is given by the Einstein relation,

$$D = \frac{k_B T}{N\zeta}. \qquad \text{Eq. 12}$$

The longest relaxation time of the Rouse chain is (see ref. (75), eq. 4.37)

$$\tau_{\mathrm{R}} = \frac{\zeta N^2 b^2}{3\pi^2 k_B T} = \frac{\langle R^2 \rangle}{3\pi^2 D}. \qquad \text{Eq. 13}$$

For the second expression, we used $\langle R^2 \rangle = Nb^2$ and Eq. 12. $\tau_{\mathrm{R}}$ is related to the vector reconfiguration time $\tau_{ij}$ of the experimentally probed segment between residues $i = 56$ and $j = 110$ of ProTα. This time is defined as a vector correlation time, $\tau_{ij} = \int_0^\infty \langle \boldsymbol{r}_{ij}(0) \cdot \boldsymbol{r}_{ij}(t) \rangle / \langle \boldsymbol{r}_{ij}^2 \rangle dt$, where $\boldsymbol{r}_{ij}$ is the vector between the positions of the two residues. Makarov found the useful relation (124)

$$\tau_{ij}/\tau_{\mathrm{R}} = \frac{\pi^2}{24} |\mu - \lambda| \{-(4 + 7\mu^2 - 12\lambda + 7\lambda^2 - 12\mu + 10\mu\lambda) + 4 - 4|\mu - \lambda| + (\mu - \lambda)^2\}, \qquad \text{Eq. 14}$$

where, in our case, $\mu = \frac{i}{N_{\mathrm{ProT\alpha}}}$ and $\lambda = \frac{j}{N_{\mathrm{ProT\alpha}}}$; with $N_{\mathrm{ProT\alpha}} = 112$, we obtain $\frac{\tau_{ij}}{\tau_{\mathrm{R}}} = 0.54$. Note that $\tau_{ij}$ is not identical to the reconfiguration time, $\tau_r$, because the latter is defined as a scalar distance correlation time, $\tau_r = \int_0^\infty \langle r_{ij}(0) \cdot r_{ij}(t) \rangle / \langle r_{ij}^2 \rangle dt$, where $r_{ij}(t) = |\boldsymbol{r}_{ij}(t)|$ is the length of the vector. We are not aware of an analytical relation between $\tau_r$ and $\tau_{ij}$. To assess the relation between the two times, we thus simulated the dynamics of Rouse chains, each for five million time steps, with different numbers of segments, $N$, ranging from 10 to 40, and with different ratios between the size of the time steps and the friction coefficients (ranging from 0.4 to 4) and found that the distance decorrelation time is smaller than the vector decorrelation time by a factor $\Xi = \frac{\tau_{\mathrm{r}}}{\tau_{\mathrm{ij}}} = 0.50 \pm 0.05$ (error is s.d. from 24 different simulations). It follows that the expected relation between the measured reconfiguration time, $\tau_r$, and the Rouse time, $\tau_{\mathrm{R}}$, is in our case:

$$\tau_r \approx \Xi \cdot 0.54 \cdot \tau_{\mathrm{R}} = 0.27 \cdot \tau_{\mathrm{R}}. \qquad \text{Eq. 15}$$

The bulk viscosity of a polymer solution (24, 75) is given by (see eq. 32 in ref. (24) and eq. 7.33 in ref. (75)):

$$\eta = \frac{c\,\zeta}{36} N b^2 + \eta_s = \frac{k_B T c_p \langle R^2 \rangle}{36\, D} + \eta_s \qquad \text{Eq. 16}$$

where $c_p = c/N$ is the concentration of protein molecules in the condensates (number of protein molecules per volume), $c$ is the concentration of chain segments (number of segments per volume) in the condensates, and $\eta_s$ is the bulk solvent viscosity (1 mPa s in our samples). The relation between reconfiguration time and viscosity is given by combining Eq. 13, Eq. 15, and Eq. 16. The error bands in Figure 4D represent the results of the models calculated from the range of experimental chain dimensions, protein concentrations, and $\Xi$.

ii) The Zimm model extends the Rouse model by including hydrodynamic interactions. It recognizes that the motion of one part of the polymer chain affects the surrounding solvent, which in turn affects the motion of other parts of the chain; the Zimm model is thus considered appropriate for polymers in dilute solution. Based on the arguments presented above, we expect our coacervates to be in the Rouse regime, but since correlation lengths and chain dimensions are in a similar range, we present the analysis based on the Zimm model for direct comparison. The Zimm model relates the center-of-mass diffusion coefficient, $D$, and the chain reconfiguration time, $\tau_Z$, to the solvent viscosity (see eq. 4.61 and 4.63 in ref. (75)), but if the viscosity in the polymer network is length scale-dependent (41), the relations can be inverted to obtain an effective solvent viscosity, $\eta_Z$, at the length scale of the chain relevant for polymer dynamics:

$$\eta_Z = \frac{8}{3\sqrt{6\pi^3}} \frac{k_B T}{\sqrt{\langle R^2 \rangle}\, D}\,, \qquad \text{Eq. 17}$$

which is equivalent to the Stokes-Einstein equation (Eq. 10) with $R_h = \frac{3\sqrt{\pi}}{8\sqrt{6}} \sqrt{\langle R^2 \rangle} \approx \frac{2}{3} R_g$,

$$\eta_Z = \frac{\sqrt{3\pi}\, k_B T \;\tau_1}{\sqrt{\langle R^2 \rangle}^3} \qquad \text{Eq. 18}$$

and $\tau_Z = \frac{8}{9\pi^2\sqrt{2}} \frac{\langle R^2 \rangle}{D}$. Eq. 19

However, in contrast to the Rouse model with entanglement (see iii), Zimm theory can only describe part of the relations we observe experimentally (Figure S7B).

iii) The Rouse model with entanglement (75, 77, 79) considers a Rouse chain diffusing in a network of other chains resulting in obstacles effectively forming tubes of diameter $a$, within which the chain can diffuse, where $a$ can also be considered the 'entanglement spacing' (78) (Figure 4C). In this picture, the center-of-mass diffusion coefficient of the chains, $D$, depends both on the friction coefficient, $\zeta$, acting on individual beads, and on the ratio between the tube diameter and the chain dimensions (see eq. 6.40 in ref. (75)),

$$D = \frac{k_B T}{3N\zeta} \frac{a^2}{\langle R^2 \rangle}\,. \qquad \text{Eq. 20}$$

The viscosity of the polymer solution can thus be expressed in terms of $D$ (see eq. 7.47, 7.43, 6.19 and 6.40 in ref. (75)) as

$$\eta = \frac{k_B T c_p \langle R^2 \rangle}{36\, D} \frac{\langle R^2 \rangle}{a^2} + \eta_s\,, \qquad \text{Eq. 21}$$

and can be used to calculate the effective tube diameter for all samples (Figure S7C), resulting in $a$ = 4 ± 2 nm (±2 nm is the variability among the samples), which is of the same order as the correlation length, $\xi$, estimated for the dense phases (125). The error bands in Figure 4D account for the slightly different entanglement spacing, chain dimensions, and concentrations in the different samples.

There are three characteristic times for chain relaxation (compare to eq. 6.106, 6.18, 6.19 in ref. (75)): $\tau_e = \frac{a^6}{3\langle R^2 \rangle^2 D}$ is the time at which the displacement of chain segments becomes comparable to the tube diameter, $a$; $\tau_{\mathrm{R_{tube}}} = \frac{a^2}{9\pi^2 D}$ is the time for chain relaxation within a tube; $\tau_{\mathrm{d}} = \frac{\langle R^2 \rangle}{3\pi^2 D}$ is the disengagement time — the time required for a chain to disengage from the tube within which it was confined. In Figure S7B, we compared the experimental chain reconfiguration time and diffusion coefficient with the three characteristic times of an entangled chain with effective tube diameter of $a$ = 4 nm and found that the chain relaxation that best describes the experimental results is $\tau_{\mathrm{d}}$. This observation suggests that the major contribution to end-to-end distance fluctuations is due to protein disengagement. Note that $\tau_{\mathrm{d}}$ and $\tau_{\mathrm{R}}$ defined for the Rouse model have the same dependence on *D*.

The viscosity of the polymer solution can also be obtained from the chain relaxation times by combining Eq. 21 with the three relations for $\tau_{\mathrm{e}}$, $\tau_{\mathrm{R_{tube}}}$, and $\tau_{\mathrm{d}}$. The finding that $\tau_r \approx \tau_d$ is valid both for the relation between chain reconfiguration time and diffusion coefficient (Eq. 4), and for the relation between chain reconfiguration time and viscosity (Eq. 5, Figure S7B). For consistency with section i), in Figure 4D and Figure S7B, we scaled $\tau_{\mathrm{e}}$, $\tau_{\mathrm{R_{tube}}}$, and $\tau_{\mathrm{d}}$ relative to $\tau_{\mathrm{r}}$ as described for $\tau_{\mathrm{R}}$ in Eq. 15.

*Molecular dynamics (MD) simulations*

All-atom explicit solvent simulations of phase-separated ProTα-H1 in 8 mM KCl as well as ProTα-protamine in 8 mM KCl and 128 mM KCl were performed using the same simulation parameters as previously described for phase-separated ProTα-H1 at 128 mM KCl (41). In brief, we employed the Amber99SBws force field (126, 127) with the TIP4P/2005s water model (128, 129). The temperature was kept constant at 295.15 K using stochastic velocity rescaling (130) ($\tau$ = 1 ps), and the pressure was kept at 1 bar with a Parrinello-Rahman barostat (131). Long-range electrostatic interactions were modeled using the particle-mesh Ewald method (132) with a grid spacing of 0.12 nm and a real-space cut-off of 0.9 nm. Dispersion interactions and short-range repulsion were described by a Lennard-Jones potential with a cutoff at 0.9 nm. Bonds involving hydrogen atoms were constrained to their equilibrium lengths using the LINCS algorithm (133). Equations of motion were integrated with the leap-frog algorithm with a time step of 2 fs, with initial velocities taken from a Maxwell-Boltzmann distribution at 295.15 K. All simulations were performed using GROMACS (134) version 2021.5. We simulated the unlabeled variant of ProTα, since the droplets under experimental conditions had 1000-fold higher concentration of unlabeled than labeled ProTα.

To obtain the starting structure of phase-separated ProTα-H1 at the desired ion concentration of 8 mM KCl, and to ensure charge neutrality, we removed 2289 potassium and 2289 chloride ions from the snapshot at 5 μs of our previous phase-separated ProTα-H1 at 128 mM KCl (41). The simulation system contained 96 ProTα and 80 H1 molecules, 129 potassium and 241 chloride ions, and 899,220 water molecules, resulting in a simulation system of 3,996,354 particles. The free production run was 3.1 μs long, with a timestep of 2 fs, employing 36 nodes (each consisting of an Intel Xeon E5-2690 v3 processor with 12 cores and an NVIDIA Tesla P100 GPU at the Swiss National Supercomputing Centre) with a performance of ~35 ns/day, corresponding to ~3 months of supercomputer time. The first 2.1 μs were treated as system equilibration and not used for the analysis.

For the ProTα-protamine simulations, we used the same all-atom force field with explicit solvent described above, but to generate an initial structure for these simulations, a phase-separated system in slab configuration (61) was obtained with coarse-grained (CG) simulations, following the strategy described previously (41). For this initial setup only, we utilized the one-bead-per-residue model that was previously developed to study the 1:1 ProTα-H1 dimer (64). Briefly, the potential energy has the following form:

$$V = \frac{1}{2}\sum_{i<N} k_b\left(d_{ij} - d_{ij}^0\right)^2 + \frac{1}{2}\sum_{i<N-1} k_\theta\left(\theta_{ijk} - \theta_{ijk}^0\right)^2 + \sum_{i<N-2}\sum_{n=1}^{4} k_{i,n}(1 + \cos(n\phi_{ijkl} - \delta_{i,n})) + \sum_{a<b} \frac{q_a q_b}{4\pi\epsilon_d\epsilon_0 d_{ab}} \exp\left[-\frac{d_{ab}}{\lambda_D}\right]$$

$$+\sum_{(a,b)} 4\varepsilon_{pp}\left(\left(\frac{\sigma_{ab}}{d_{ab}}\right)^{12}-\left(\frac{\sigma_{ab}}{d_{ab}}\right)^{6}\right),\qquad \text{Eq. 22}$$

where $i, j, k, l$ denote consecutive residues. The first term represents the harmonic bond energy with force constant $k_b = 3.16 \times 10^5$ kJ.mol$^{-1}$nm$^{-2}$, and the second term represents the angle energy with force constant $k_\theta = 6.33 \times 10^2$ kJ mol$^{-1}$rad$^{-2}$; reference values for $d_{ij}^0$ and $\theta_{ijk}^0$ were taken from an extended backbone structure (0.38 nm; $2\pi/3$ rad). The third term represents a sequence-based statistical torsion potential taken from the Go model of Karanicolas and Brooks (135), which was applied to all residues. The fourth term represents a screened coulomb potential, with Debye screening length $\lambda_D$, applied to all residues with non-zero charges $q_i$; $\epsilon_0$ is the permittivity of free space; the dielectric constant, $\epsilon_d$, was set to 80. The fifth term represents a generic short-range attractive potential applied to all residue pairs. This interaction is characterized by a contact distance $\sigma_{ab} = (\sigma_a + \sigma_b)/2$, where $\sigma_{a,b}$ are the residue diameters (all 0.6 nm) determined from residue volumes (136), and a contact energy $\varepsilon_{pp}$, which is the same for all residue pairs and was set to $0.16\ k_B T$, or $\sim 0.40$ kJ/mol$^{-1}$. The Debye length, $\lambda_D$, is given by

$$\lambda_D = \left(\frac{\epsilon_d \epsilon_0 k_B T}{2e^2 I}\right)^{1/2},\qquad \text{Eq. 23}$$

where $k_B$ is the Boltzmann constant, *T* the temperature, *e* the elementary charge, and *I* the ionic strength.

Using this coarse-grained model, 96 ProTα and 197 protamine molecules (1:1 charge balance) in an initially extended configuration were placed on a rectangular grid in a 60-nm cubic box, and the energy of the system was minimized with the steepest-descent algorithm. The system was further relaxed in a short NVT run at 500 K and an implicit ionic strength of 500 mM. In the next step, the box edge was decreased to 22.41 nm in a 56-ps NPT run with reference pressure set to 20 bar to obtain an average protein density close to that of the dense phase in experiment. The system configuration was further randomized via a 1.5-μs NVT run (using a 10-fs time step) at 500 K and an implicit ionic strength of 500 mM to ensure relatively uniform protein density in the box. Each chain from the final CG structure was independently reconstructed in all-atom form using a lookup table from fragments drawn from the PDB, as implemented in Pulchra (136). Side-chain clashes in the all-atom representation were eliminated via a short Monte Carlo simulation with CAMPARI (137) using the ABSINTH energy function (137), in which only the side chains were allowed to move. Due to the large number of proteins and the relatively high density of proteins in the box, the first CAMPARI moves were performed using a soft-core Lennard-Jones (LJ) potential, which has an energy cap, thus avoiding the very large energies resulting from the exclusion in the first Monte Carlo moves at the beginning of the CAMPARI simulation. Subsequently, the soft-core LJ potential was gradually transformed into a global soft-core form by increasing the CAMPARI parameters FMCSC_FEG_IPP and FMCSC_FEG_ATTLJ. The following pairs of FMCSC_FEG_IPP and FMCSC_FEG_ATTLJ values were used: 0.5 and 0.35, 0.7 and 0.6, 0.9 and 0.85, 0.95 and 0.92, 0.98 and 0.96, 0.99 and 0.97, 0.9925 and 0.975, 0.995 and 0.99. Taking the relaxed configuration obtained with CAMPARI, the box edge was extended to 41 nm in the Z direction, and the resulting system was set up with the all-atom Amber ff99sbws protein force field (126, 127) in GROMACS and energy-minimized using the steepest-descent algorithm. To eliminate any non-proline cis bonds that might have emerged during all-atom reconstruction, we ran a short simulation in vacuum with the dielectric constant set to 80 with periodic boundaries, using a version of the force field that strongly favors trans peptide bonds (61) and applying weak position restraints to the protein backbone atoms and dihedral angles (5 kJ/mol/rad).

Subsequently, the simulation box was filled with TIP4P/2005s water[44] and energy-minimized. Up to this point, the same setup was used for both 8 mM and 128 mM ProTα-protamine simulations. In the next step, 1587 potassium and 1596 chloride ions were added to the simulation box for ProTα-protamine at 128 mM KCl (2,612,851 particles in total), and 99 potassium and 108 chloride ions were added to the simulation box for ProTα-protamine at 8 mM KCl (2,621,779 particles in total), to match the ionic strength of the buffer used in the experiments and to ensure charge neutrality. In the next step, both systems were again energy-minimized, and a 10-ns MD run was performed with strong position restraints on protein backbone atoms ($10^5$ kJ mol$^{-1}$ nm$^{-2}$) to stabilize the trans isomer for any peptide bonds that had isomerized in the previous step. The final structures of these 10-ns runs with backbone restraints were used for the production runs (without restraints), using GROMACS (134) versions 2021.5 and 2024.1. The free production runs at 128 mM and 8 mM KCl were 4 μs long, with a timestep of 2 fs. For the first 2.5 μs of both simulations, we employed 48 nodes (each node consisting of an Intel Xeon E5-2690 v3 processor with 12 cores and an NVIDIA Tesla P100 GPU at the Swiss National Supercomputing Centre) with a performance of ~63 ns/day, corresponding to ~40 days of supercomputer time. The last 1.5 μs of both simulations were run on NVIDIA Grace-Hopper (GH200) nodes at the Swiss National Supercomputing Centre, using 192 CPUs and 4 GPUs per simulation, with a performance of ~62 ns/day, corresponding to ~24 days of supercomputer time. The first 3 μs of both simulations were treated as system equilibration (given the change in protein density, see Fig. S4B) and not used for analysis.

*Analysis of MD simulations*

For chain dynamics with relaxation times much shorter than the total simulation time, distance correlation functions can be calculated directly from the simulations and compared with experiment, as demonstrated previously for complex coacervates of H1 + ProTα (41). For the simulations of protamine + ProTα, however, chain relaxation is too slow for this approach to be reliable. To estimate long-timescale dynamics and correlation functions from the condensate trajectories also for these cases, the dynamics of each ProTα chain were described as one-dimensional diffusion along the coordinate defined by the separation in space, *r*, of the residues that were dye-labeled in experiment (here we use the distance between Cα atoms of residues 58 and 112) (Figure S9). That is, the separation is considered to diffuse on a free energy surface, *F*(*r*), with a diffusion coefficient $D_F$ (more generally, position-dependent $D_F(r)$), whose parameters we determine from the simulations. This distance coordinate is first discretized into *b* bins of equal width, from which the number of transitions, $N_{ji}(\Delta t)$, from bin *i* to bin *j* after a lag $\Delta t$ during the simulations is counted. These statistics are combined from all ProTα chains, considering that they are expected to be indistinguishable. Discretized free energies and diffusion coefficients were optimized via Monte Carlo simulations using the likelihood function

$$\ln L = \sum_{i,j} N_{ji}(\Delta t) \ln p(j, t + \Delta t | i, t), \quad \text{Eq. 24}$$

where the propagators $p(j, t + \Delta t | i, t)$ describing the conditional probability of being in bin *j* at time $\Delta t$ after having been in bin *i* are obtained from the discretized diffusion model as previously described (138, 139): in short, the discretized dynamics is mapped to a chemical kinetics scheme describing evolution of populations in the bins, $\dot{\mathbf{P}}(t) = \mathbf{KP}(t)$, where $\mathbf{P}(t)$ is the vector of the bin populations at time *t*, and **K** is a rate matrix derived from the diffusion coefficient(s) $D_F$ (or $D_{F,i}$ for position-dependent $D_F$) and free energies $F_i$ associated with each bin according to the scheme of Bicout and Szabo (138, 140). The propagators are then given by $p(j, t + \Delta t | i, t) = (\exp[\Delta t \mathbf{K}])_{ji}$. In estimating the most probable parameters from the data, a uniform prior is used for the diffusivities and free energies. The statistical error on the derived parameters is determined by generating synthetic data sets with the same number *M* of individual distance trajectories *r*(*t*) as the original, specifically by choosing *M* trajectories randomly with replacement, and refitting the model. The error is taken as the standard deviation of the parameters across all synthetic data sets (141).

We can compute the normalized correlation functions directly from the discretized diffusion model via (142)

$$C(t) = \frac{\sum_{n=2}^{b} (\mathbf{r} \cdot \boldsymbol{\psi}_n^R)^2 \exp[\lambda_n t]}{\sum_{n=2}^{b} (\mathbf{r} \cdot \boldsymbol{\psi}_n^R)^2}, \quad \text{Eq. 25}$$

where the elements of **r** are the centers of each bin on the distance coordinate, $\boldsymbol{\psi}_n^R$ is the *n*th right eigenvector of **K** ($\boldsymbol{\psi}_1^R$ is the stationary eigenvector), and $\lambda_n$ is the *n*th eigenvalue. Similarly, the correlation times are given by

$$\tau_c = -\frac{\sum_{n=2}^{b}(\mathbf{r}\cdot\boldsymbol{\psi}_n^R)^2\lambda_n^{-1}}{\sum_{n=2}^{b}(\mathbf{r}\cdot\boldsymbol{\psi}_n^R)^2}. \qquad \text{Eq. 26}$$

Errors in correlation functions and correlation times were estimated using the same procedure as for the diffusion model parameters. In our application of the method to the condensate trajectories, we have used 30 equal-width bins between 2 and 10 nm, and a lag time of 200 ns.

A key assumption of this method is that the dynamics is, indeed, well approximated as diffusive after the chosen lag time $\Delta t$. If this is true, then the model should become independent of lag time beyond this point. To assess this effect, we computed the correlation time as a function of the lag time and observed that after lag times of around 200 ns it appears to be converging toward a limiting value (Figure S9D). One challenge for using even longer lag times is the limited length of the simulations, resulting in insufficient statistically independent observations. Separately from the statistical error estimate described above, we also estimated the systematic error associated with the choice of lag time by using correlation times computed at 100 and 300 ns lag times as lower and upper error bars, respectively. A second assumption we have made is that the diffusion coefficient should be uniform, i.e. not dependent on the position on the distance coordinate. This was motivated by our finding that using an explicitly position-dependent diffusion coefficient resulted in very little position dependence, as demonstrated in Figure S9B. Although this conclusion differs from some earlier work (143), this is most likely because we do not significantly sample the very short distances where the position dependence of $D_F$ emerged in that study.

The average number of H1 or protamine molecules that simultaneously interact with a single ProTα chain, as well as the average number of ProTα chains that simultaneously interact with a single H1/protamine molecule (Fig. 3d) in the dense-phase simulation were quantified by calculating the minimum distance between each ProTα and each H1/protamine for each simulation snapshot. The two molecules were considered to be in contact if the minimum distance between any two of their Cα atoms was within 1 nm. Distances between Cα atoms were used instead of the commonly used distances between all atoms of the residues to facilitate the large calculations. The 1-nm cutoff between the Cα atoms of two residues yields similar results as the commonly used 0.6-nm cutoff for interactions between any pair atoms from the two residues (134). The same contact definition was employed when calculating residue-residue contacts (Fig. 3e): Two residues were considered to be in contact if the distance between their Cα atoms was within 1 nm.

Lifetimes of residue-residue contacts were calculated by a transition-based or core-state approach (41, 144). For each pair of residues, a contact was based on the shortest distance between any pair of heavy atoms, one from each residue. Starting from an unformed contact, contact formation was defined to occur when this distance dropped below 0.38 nm; an existing contact was considered to remain formed until the distance increased to more than 0.8 nm (144). Average lifetimes of each residue-residue contact were calculated by dividing the total bound time by the total number of contact breaking events for that contact. Intra-chain contacts were not included in the analysis. Average lifetimes of each pair of ProTα-H1 and ProTα-protamine residues (averaged over the different combinations of ProTα and H1/protamine chains that the two residues could be part of) were calculated by dividing the total contact time (summed over all combinations of ProTα and H1/protamine chains) of a specific residue pair by the total number of the contact breaking events for the same residues (summed over the same combinations of chains). Similarly, to calculate average lifetimes of residue-residue contacts according to the residue type, we first identified all contacts involving a particular pair of residue types, in which one residue was from the ProTα chain and the second was from either H1/protamine or ProTα. Subsequently, the average lifetime of that residue-residue combination was calculated by dividing the total bound time by the total number of contact breaking events for the contacts involving those residue types. Excess populations of contacts between specific types of residues were determined by dividing the average number of observed contacts for a pair of residue types by the value that would be expected if residues paired randomly in a mean-field approximation. The average number of contacts for a pair of residue types was calculated as a sum of all times that residues of those types were in contact, divided by the simulation length. The expected average number of contacts between two residue types (type 1 and 2) was calculated as *N f(1) f(2),* where *N* is the average total number of contacts, and *f(1)* and *f(2)* are the fraction of residues of type 1 and 2, respectively.

The mean squared displacement (MSD) of individual residues within ProTα molecules were calculated as a function of delay time using the Gromacs function *gmx msd*. MSD curves of each ProTα residue for each of the 96 chains in simulations with protamines as well as in simulation with H1 at 8 mM KCl were calculated from the last microsecond of each of the simulations, using residue coordinates every 100 ps. MSD curves of each ProTα residue for each of the 96 chains in ProTα-H1 simulation at 128 mM KCl were calculated previously in four 1-µs blocks, using residue coordinates every 100 ps.

All of these analyses are consistent with the previous reports in Galvanetto *et al*. (41).

Distance distributions of the closest and second-closest residues (Figure 3H) were computed between all residues of all polypeptides in the simulations. The distances were considered between specific atoms in the side chains: Cδ for glutamate; Cγ for aspartate; Nζ for lysine; Cζ for arginine; and Cβ for alanine. The distributions shown in Figure 3H are averages of 100 structures taken every 1 ns. The percentages of contacts that are exchanged, broken and reformed, or remain unbroken (Figure 3K) were determined by first identifying the lysines in H1 and the glutamates in ProTα in the H1–ProTα slab at 128 mM KCl that form a close contact at time $t_0$, defined as being within 0.43 nm, corresponding to the sharp peak in the distribution of closest lysine contacts (Figure 3H, top panel). These contacts were then tracked for 100 ns and the fraction of contacts that separated beyond 0.43 nm during this time was quantified, at which point another lysine forms a closer contact. Additionally, we quantified the fraction of contacts that remain intact for the entire 100 ns and the fraction that, while not continuously maintained, breaks and reforms between the same two residues. The same analysis was performed for arginine (instead of lysine) in the protamine–ProTα dense phase at 128 mM KCl, using a distance threshold of 0.53 nm, corresponding to the peak in the distribution of closest arginine contacts (Figure 3H, middle panel).

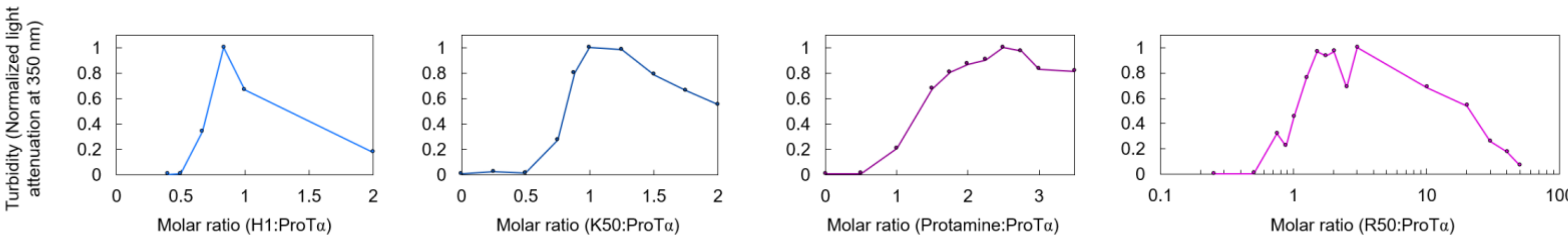

**Figure S1. Phase separation is most pronounced for protein mixtures near charge balance.** The extent of droplet formation was assessed using turbidity at a constant concentration of 10 μM ProTα and varying amounts of its polycationic partners at 120 mM KCl. Maximal phase separation was observed close to molar ratios, λ, where the charges of the two polymers balance (net charges: ProTα, -44; H1, +53; K50, +50; protamine, +22; R50, +50). From left to right, $\lambda_{H1:ProT\alpha}$ = 44:53 ≈ 0.8:1, $\lambda_{K50:ProT\alpha}$ = 44:50 ≈ 0.9:1, $\lambda_{protamine:ProT\alpha}$ = 44:22 = 2:1, $\lambda_{R50:ProT\alpha}$ = 44:50 ≈ 0.9:1. We note that the arginine-rich samples (protamine and R50) tend to phase-separate even with an excess of the polycationic partner, reflecting the complex interactions of arginine beyond Coulomb interactions (51, 145).

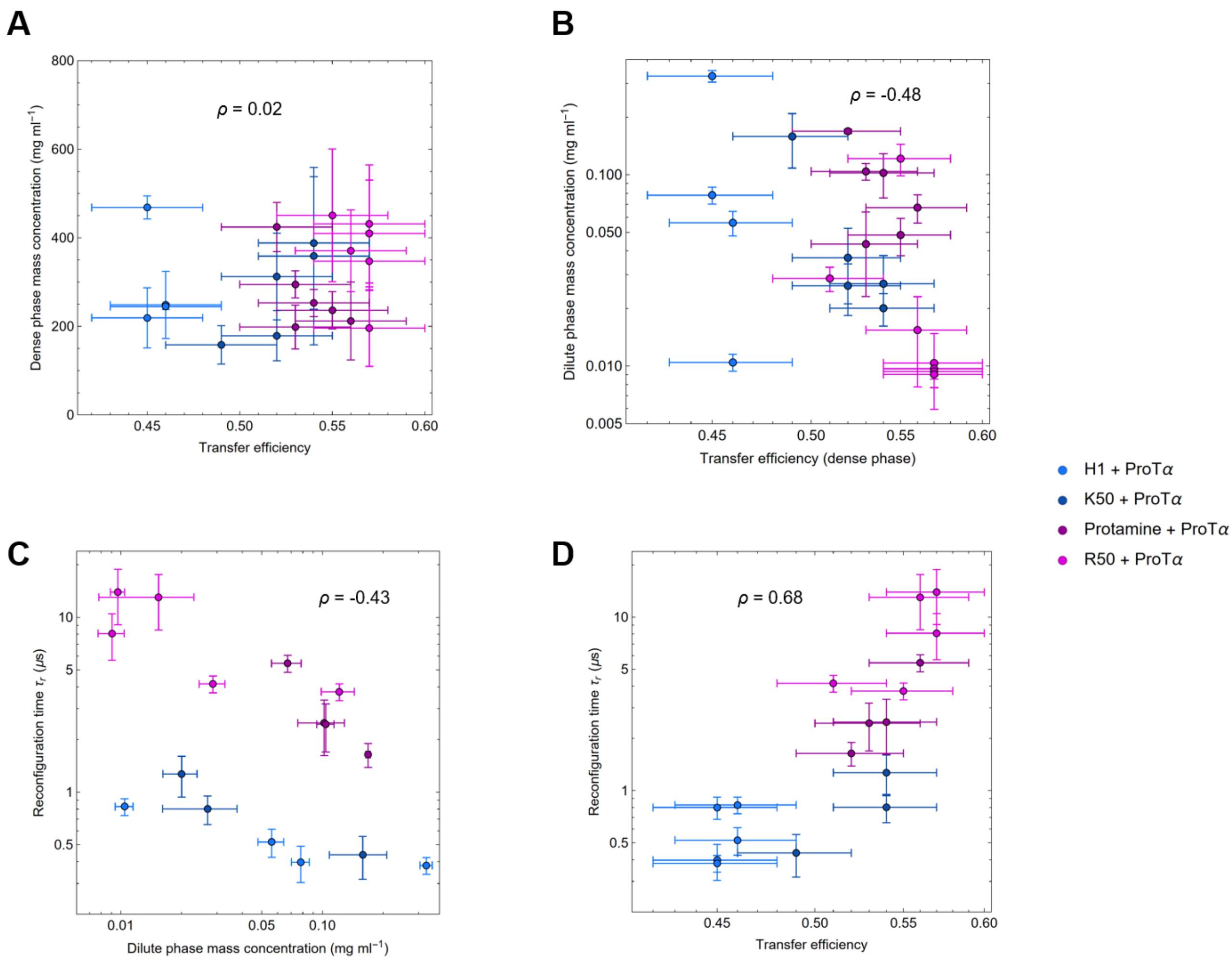

**Figure S2. Correlations between thermodynamic and dynamic quantities in the dense and dilute phases.** (**A**) ProTα transfer efficiency *vs* dense phase protein concentration shows no correlation. (**B**) Dense-phase ProTα transfer efficiency *vs* dilute-phase protein concentration shows a slight anticorrelation, indicating that both thermodynamic quantities are proxies for the interaction strength between polymers: the stronger the interaction, the lower the dilute phase concentration and the smaller the chain dimensions (82). (**C**) Correlations between chain reconfiguration time, $\tau_r$, and dilute phase mass concentration, and (**D**) between $\tau_r$ and transfer efficiency indicate relations between the molecular dynamics within condensates and the intermolecular interactions of the systems, as proposed by An *et al*. (32). $\rho$ are the Pearson correlation coefficients.

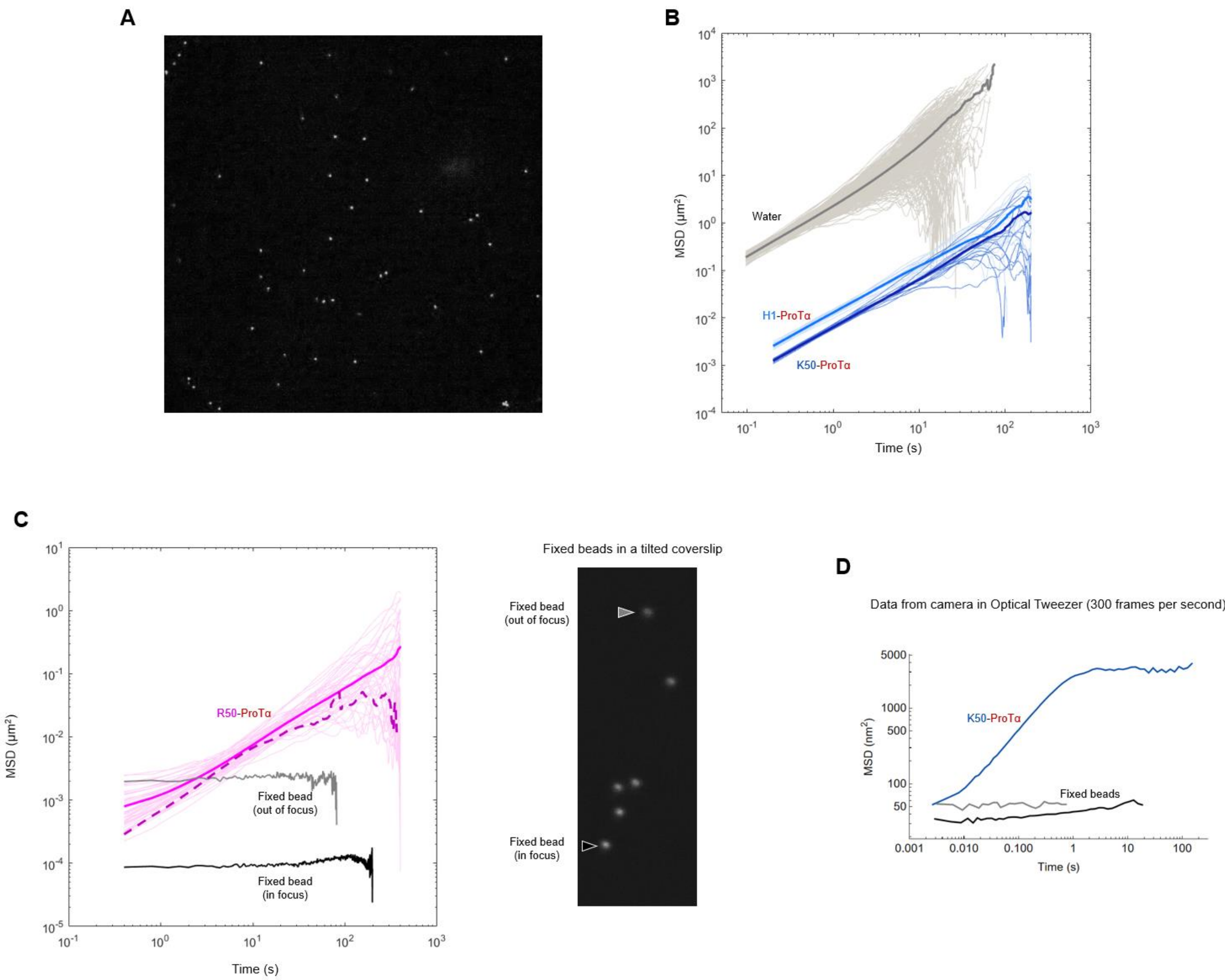

**Figure S3. Passive microrheology shows Brownian diffusion of beads in the dense phase down to the millisecond timescale.** (**A**) Example of a fluorescence micrograph of 500-nm beads in a K50-ProTα droplet (image size: 150x150 μm). (**B**) Mean-squared displacement (MSD) of individual 500-nm beads and their averages (solid thick lines) in water, in K50-ProTα dense phase in 60 mM KCl, and in H1-ProTα dense phase in 120 mM KCl, show Brownian diffusion. (**C**) MSD of 500-nm beads in R50-ProTα dense phase in 90 mM KCl (right) apparently deviates from Brownian diffusion at short times, which might be mistaken to suggest the approach of the elastic plateau (80). Further examination of the MSD of beads fixed on a cover slide at different positions relative to the focal plane (left) indicates that this deviation is an artifact due to the limited precision in determining the position of the beads. A stricter threshold for automatic bead identification can reduce this artifact (dashed line), but it also reduces the number and length of individual trajectories, making viscosity determination problematic over longer time periods. (**D**) The MSD as a function of time for a single polystyrene bead (1 μm diameter) trapped in the dense phase of K50-ProTα with optical tweezers also appears to change slope at short times. Similarly, the deviation observed for MSD values <100 $nm^2$ is likely to be caused by the limited precision in determining the position of the beads. (Tracking was performed from brightfield images taken with maximum LED illumination, see Methods.)

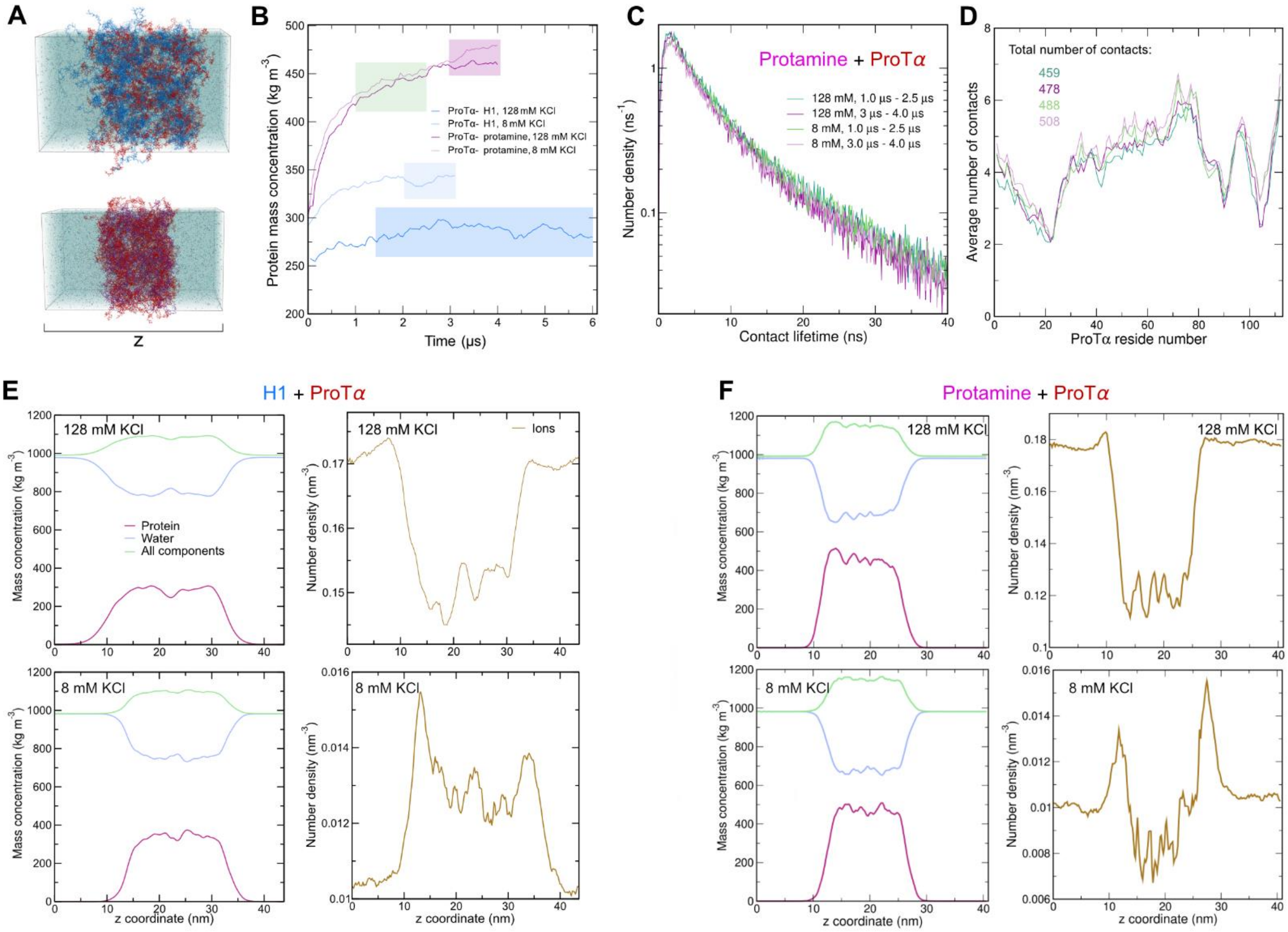

**Figure S4. Mass concentrations and ion distributions in MD simulations.** (**A**) Illustrations of H1–ProTα and protamine–ProTα simulations in slab geometries. Densities were calculated along the z-axis. (**B**) Protein concentration in the dense phases as a function of time, calculated in 50-ns blocks. The parts of the simulations where the protein density strongly increases with time were treated as equilibration and omitted from the analysis. The parts of the simulations that were analyzed are highlighted by the shaded boxes. (**C**-**D**) The effect of protein density differences on the contact lifetimes (E) and on the number of contacts per residue (F) is small. Two segments of the protamine–ProTα trajectories were analyzed, spanning 1 to 2.5 μs and 3 to 4 μs, respectively (indicated by the green and purple boxes in panel B). Despite the difference in protein density of ~6% between the two time windows, their lifetime distributions are very similar, while the average contact lifetimes increase slightly (by ~10%), mostly due to a relatively small number of very long contacts. The total number of contacts between ProTα and protamine (D) increases by ~4%, consistent with the slight increase in protein density. **(E–F)** Mass concentrations of protein, water, all components (protein, water, and ions; left panels), and number density of ions (right panels) along the z-axis of the simulation box in the four different simulations. The water density in the protamine–ProTα simulations is lower than the water density in the H1–ProTα simulations, which is consistent with the higher protein density observed in the protamine–ProTα simulations. For the simulations with a total salt concentration of 128 mM KCl, ion concentrations within the dense phase are decreased relative to the dilute phase in both the H1–ProTα and protamine–ProTα

simulations, in parallel with the decrease in water content within the dense phases. Conversely, for the simulations with a total salt concentration of 8 mM KCl, ion concentrations are increased in the dense phase relative to the dilute phase in both the H1–ProTα and protamine–ProTα simulations. It is worth noting that a dependence of salt partitioning on the salt concentration has also been observed for synthetic complex coacervates (146).

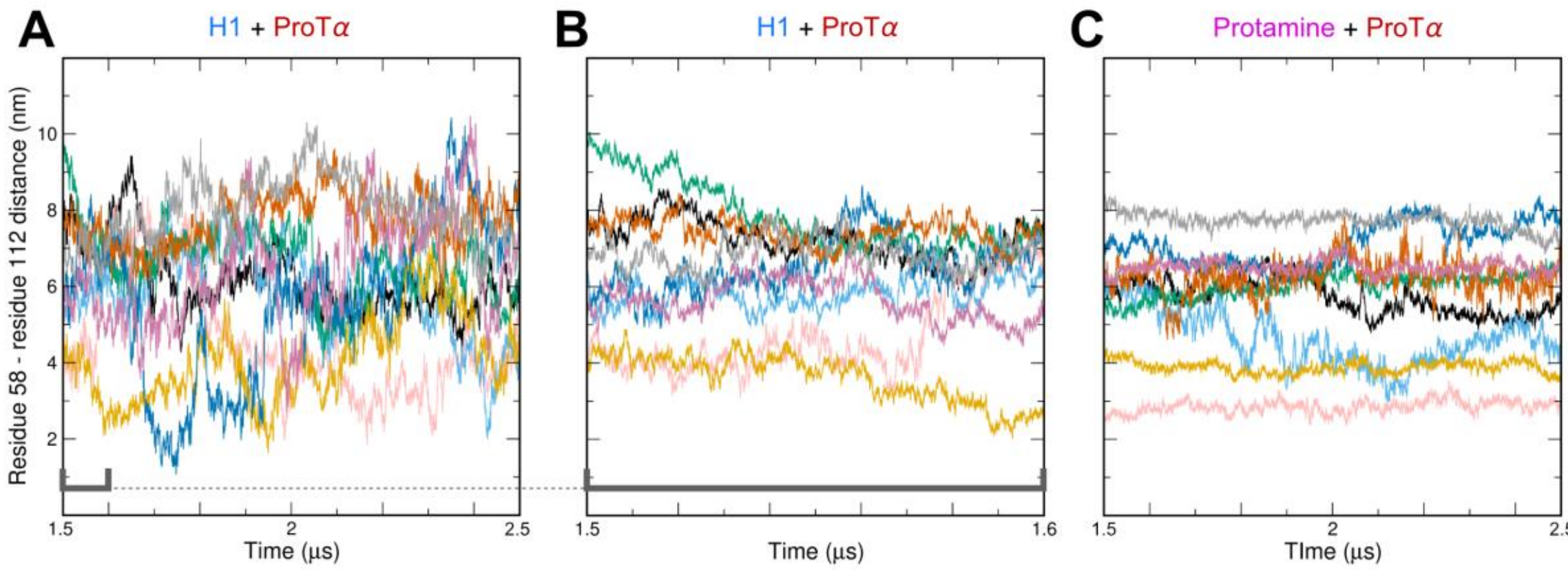

**Figure S5. Comparison of ProTα chain dynamics in H1–ProTα and protamine–ProTα dense phase at 128 mM KCl from MD simulations.** (**A**) Examples of intrachain distance fluctuations between residues 58 and 112 for 9 of the 96 ProTα chains (chain 10, 20, ... , 90) in the H1–ProTα dense phase during 1 µs. (**B**) Same as **A**, but illustrated for the first 0.1 µs. (**C**) Examples of intrachain distance fluctuations between residues 58 and 112 for 9 of the 96 ProTα chains in the protamine–ProTα dense phase over 1 µs are qualitatively comparable to the distance fluctuations in the H1–ProTα slab during 0.1 µs, illustrating the ~10-fold slower chain dynamics in the protamine–ProTα dense phase, which is also reflected in the chain reconfiguration times (Figures 3 and S9).

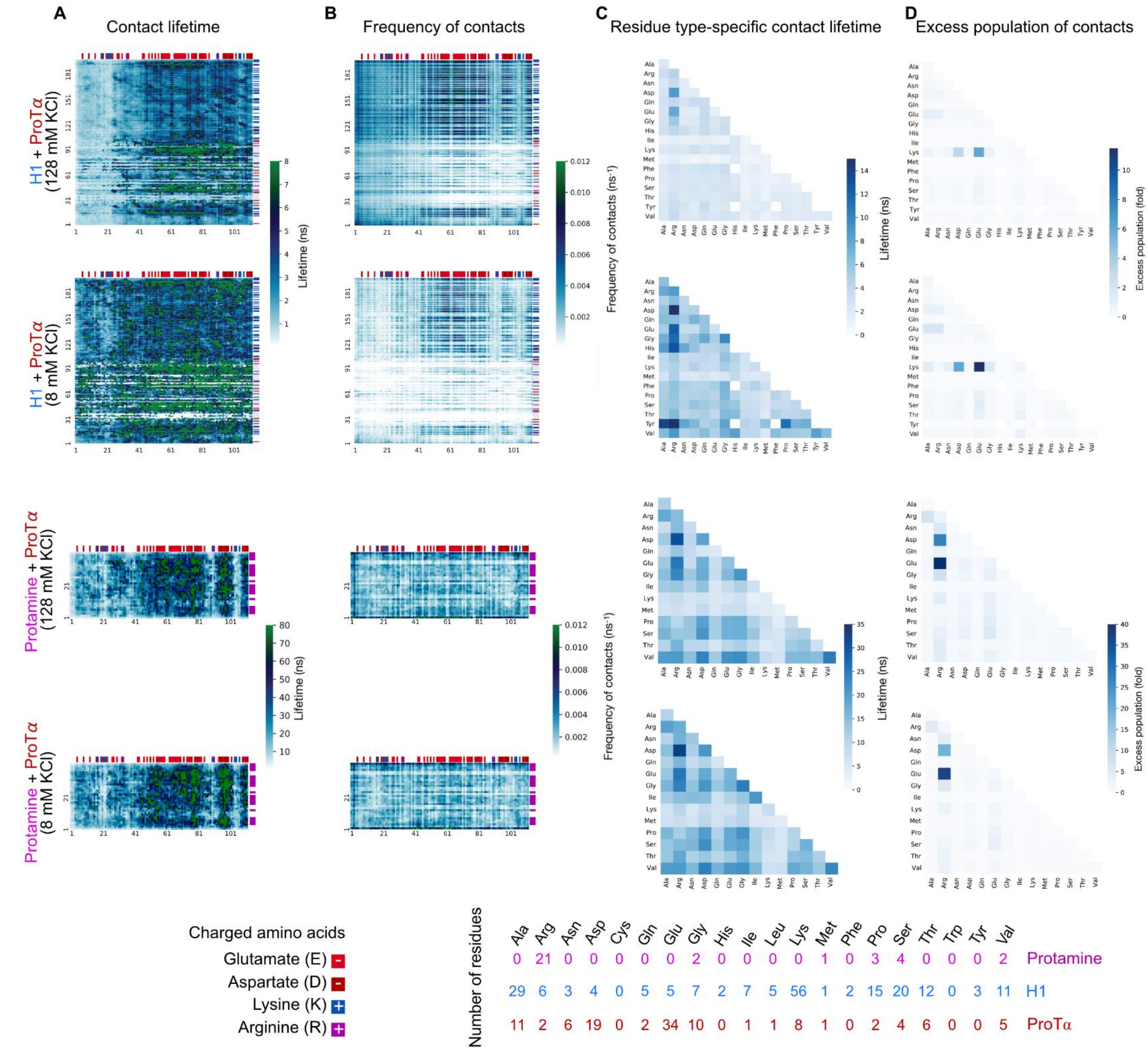

**Figure S6. Contact lifetime heatmaps from MD simulations.** (**A**) Average lifetime of residue-residue contacts, calculated by considering all instances of these residues across all chains. Numbers on the bottom and left of the heatmaps denote the residue numbers of ProTα and H1 or protamine, respectively. Note that H1 contains a folded globular domain (GD) between residues 22 and 96. Bars at the top and right of the plots denote charged residues (color code as in Figure 1A). (**B**) Frequency of contacts (defined as the number of new contacts made by one ProTα residue per nanosecond). Panels **A** and **B** indicate that: (i) decreasing salt concentration increases the lifetime of residue-residue contacts (see also Figure 3F); (ii) contact times in arginine-rich droplets (Protamine–ProTα) are longer than in lysine-rich droplets (H1–ProTα) (see also Figure 3F); (iii)

charge-charge contacts are the most frequent but the most short-lived contacts in the lysine-rich condensates, whereas charge-charge contacts are both the most frequent and the longest-lived in the arginine-rich condensates, reflecting the propensity of arginine to form multivalent contacts (see also Figure 3J). (**C**) Average lifetimes of residue-residue contacts classified by residue type. Residue pairs that are never observed (white squares) and extremely long-lived pairs (dark blue) typically correspond to residue types that are rare in the ProTα and H1/protamine sequences. (**D**) Excess populations of contacts between specific types of residues (determined by dividing the average number of observed contacts for a pair of residue types by the value that would be expected if residues paired randomly in a mean-field approximation, see Methods). The large excess of contacts between charged residues indicates that their interactions are the most favorable in both lysine- and arginine-rich condensates. Although these contacts are the most frequent, their lifetimes are in the same range as those of other residue pairs (see **C**). In addition, the excess of charged residue interactions is more pronounced in arginine-rich condensates than in lysine-rich condensates, in line with the propensity of arginine to form multivalent contacts (50, 62, 147, 148).

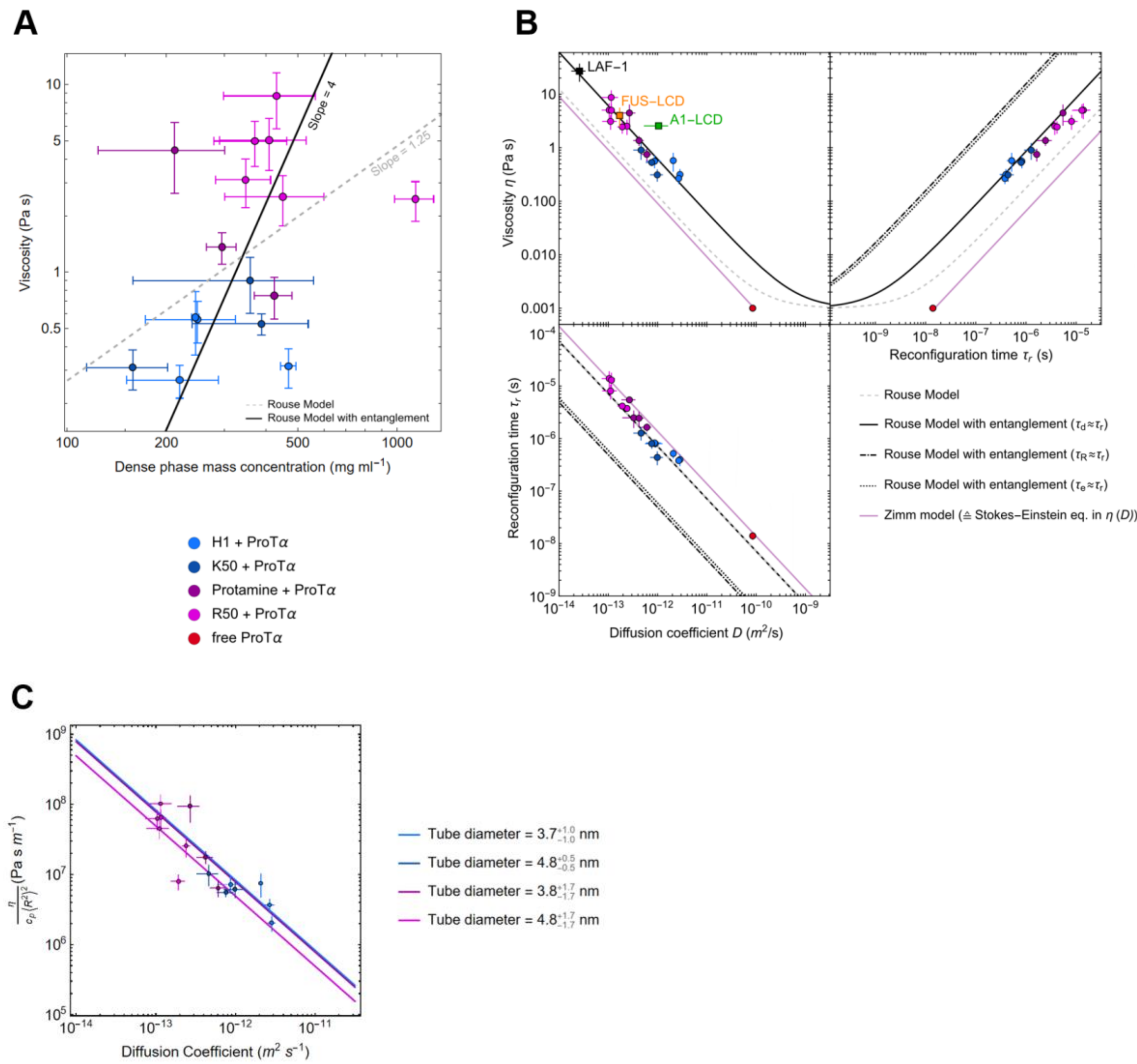

**Figure S7. Comparison of experimental data with different polymer models.** (**A**) The dependence of viscosity, $\eta$, on the dense-phase concentration, $c_p$, is better described by the Rouse theory with entanglement, which predicts a viscosity dependence between $\eta \propto {c_p}^{3.75}$ (see equation 7.65 in ref. (75)) and $\eta \propto {c_p}^{4.25}$ (149) (total least squares = 2.7 for a slope of 4), rather than by the Rouse theory without entanglement, which predicts $\eta \propto {c_p}^{1.25}$ (149) (total least squares = 6.7 for a slope of 1.25). (**B**) Comparison of the experimental viscosity, diffusion coefficient, and chain reconfiguration time with the predictions of the Rouse model, the Zimm model, and the Rouse model with entanglement from Equations 1-5 and 13-21 (see Methods). The viscosities and diffusion coefficients of LAF-1 (59), A1-LCD (33, 82), and FUS-LCD (83, 84) are from previous reports. The average tube diameter used in the Rouse model with entanglement is obtained as shown in **C** ($a$ = 4±2 nm). The Rouse model with entanglement predicts three different chain relaxation times: $\tau_\mathrm{e} = \frac{a^6}{3\langle R^2\rangle^2 D}$ is the time at which the displacement of chain segments becomes comparable to the effective tube diameter, $a$; $\tau_{\mathrm{R_{tube}}} = \frac{a^2}{9\pi^2 D}$ is the time for chain relaxation within a tube; $\tau_\mathrm{d} = \frac{\langle R^2\rangle}{3\pi^2 D}$ is the disengagement time — the time required for a chain to disengage from the tube within it was confined. Since it is not obvious which of these relaxation times corresponds to the experimental reconfiguration time $\tau_\mathrm{r}$ (i.e. the measurement of the end-to-end fluctuations of single chains with FRET-based nsFCS), we compare the three models (black lines). The chain relaxation that best describes the experimental results is $\tau_{\mathrm{d}}$, suggesting that the experimentally observed end-to-end distance fluctuations are dominated by chain disengagement. (**C**) The reduced viscosity, $\frac{\eta}{c_p \langle R^2 \rangle^2}$, allows us to evaluate the relation between viscosity and diffusion coefficient *D* to obtain the best estimate for the tube diameter, $a$ (see Eq. 4), taking into account the contribution from the slightly different chain dimensions, $\langle R^2 \rangle$, and protein concentrations, $c_p$, in the individual samples. The solid lines correspond to the relation $\frac{\eta}{c_p \langle R^2 \rangle^2} = \frac{k_B T}{36\, D} \frac{1}{a^2} + \frac{\eta_s}{c_p \langle R^2 \rangle^2}$, where the tube diameters, $a$, were calculated for each individual sample as $a = \sqrt{\frac{k_B T}{36\, D} \frac{c_p \langle R^2 \rangle^2}{\eta}}$. The mean values and standard deviations of $a$ for the individual samples are reported in the legend. For the comparison with other models, we used the average value $a$ = 4±2 nm, which is in agreement with the value independently estimated using only $c_p$ and $\langle R^2 \rangle$ following ref. (81).

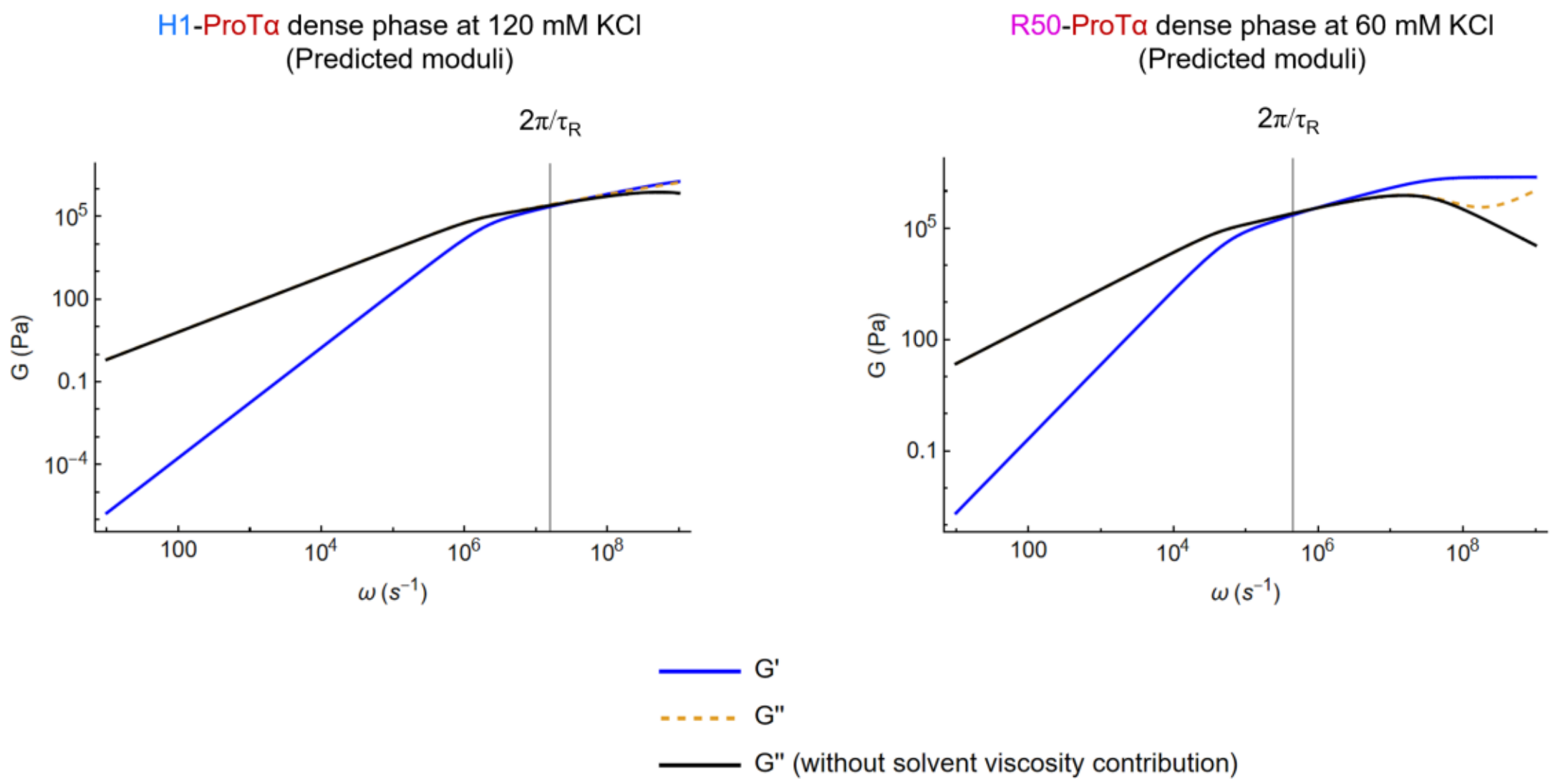

**Figure S8. Theoretical storage and loss moduli of the samples with the fastest (left) and slowest (right) chain reconfiguration time, $\tau_r$.** The Rouse model (24) can be used to calculate the storage modulus $G'(\omega) = c_p k_B T \sum_p \frac{\omega^2 {\tau_p}^2}{1+\omega^2 {\tau_p}^2}$ and the loss modulus $G''(\omega) = \omega \eta_s + c_p k_B T \sum_p \frac{\omega \tau_p}{1+\omega^2 {\tau_p}^2}$ of a polymer solution with a concentration of chains, $c_p$, and solvent viscosity, $\eta_s$, where $\omega$ is the angular frequency, and $\tau_p$ is the relaxation time of the *p*-th Rouse mode ($\tau_p = \frac{\tau_R}{p^2}$, where the longest relaxation mode, $\tau_1 \triangleq \tau_R$, is related to the reconfiguration time of the experimentally observed chain segment by the Makarov relation (124), $\tau_R = \frac{\tau_r}{0.27}$, see Methods). This theory has previously been shown to accurately describe experimental data on synthetic polymers (150). We note that even in the sample with the slowest reconfiguration time investigated here (R50-ProTα at 60 mM KCl), the predicted crossover frequency, $\frac{2\pi}{\tau_R}$, is in the megahertz range, which is difficult to access using conventional rheological methods (20). Similarly, in the case of Rouse theory with entanglement, the crossover frequency is expected to be the inverse of the disengagement time, $\tau_d$, (75, 80) which corresponds to the experimental chain reconfiguration time, $\tau_r$ (see Figure S7B).

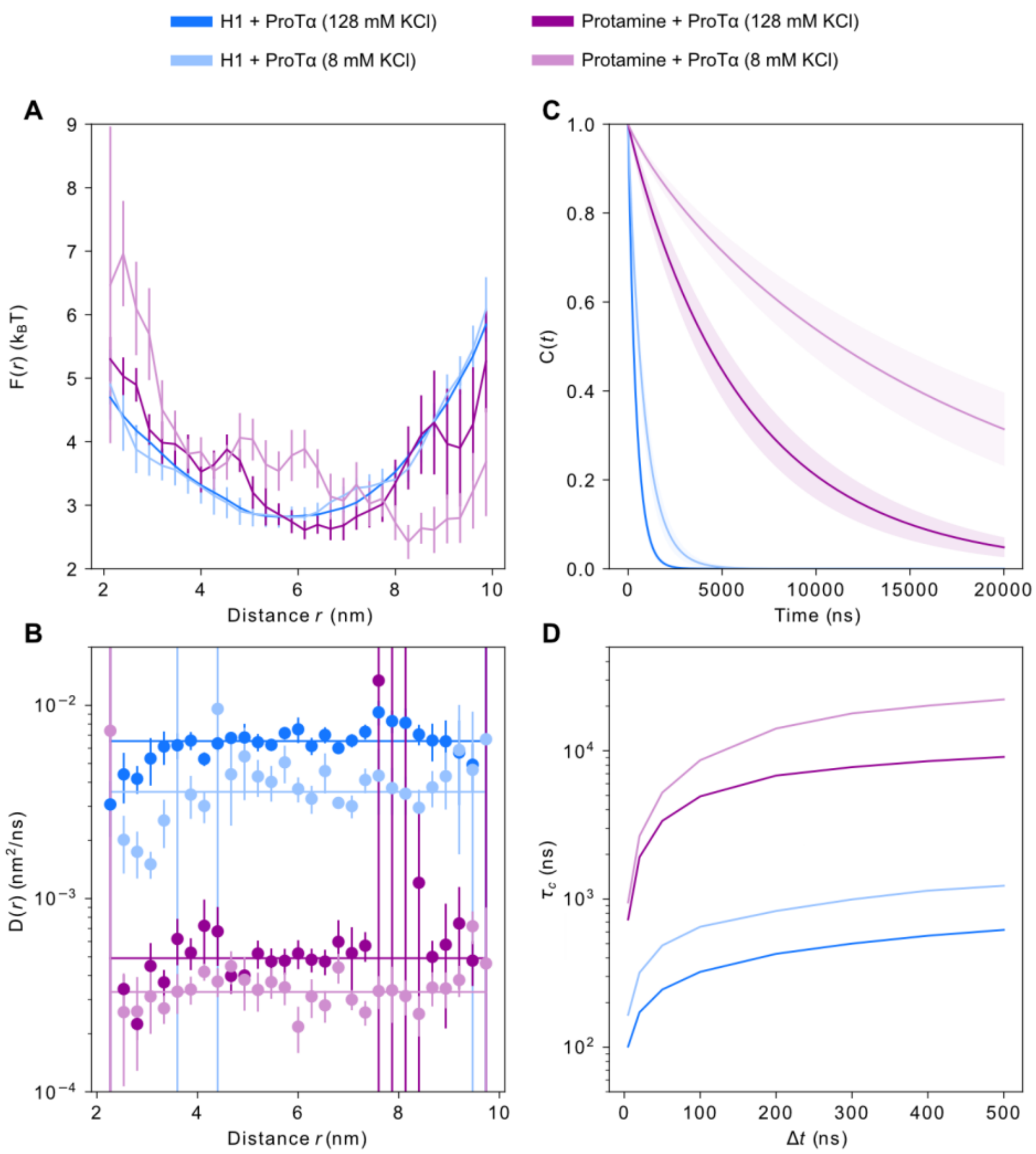

**Figure S9. Computing distance decorrelation times from simulation using a diffusion model.** For each system, the dynamics of the 58-112 distance is modeled using a discretized diffusion model described by (**A**) free energies, *F*(*r*), and (**B**) diffusion coefficients, *D*(*r*), determined from the molecular simulations. Allowing position-dependent diffusion coefficients (symbols in (**B**)) shows that constant *D* (horizontal lines in (**B**)) is a good approximation. The model allows properties beyond the ~1 μs time scale of the equilibrated portion of most of the simulations to be estimated, such as (**C**) correlation functions and (**D**) correlation times. Based on the convergence of correlation time, $\tau_c$, with lag time, $\Delta t$, in (**D**), a lag time of 200 ns was chosen for all systems and was used in the models for (**A**)-(**C**). Note that for the more rapidly relaxing H1+ProTα condensates, the distance correlations can be calculated directly from the simulation trajectories without using a diffusion model and yield similar values.

| | |
|---|---|
| ProTα (unlabeled) | GPMSDAAVDTSSEITTKDLKEKKEVVEEAENGRDAPANGNANEENGEQEA DNEVDEEEEEGGEEEEEEEEGDGEEEDGDEDEEAESATGKRAAEDDEDD DVDTKKQKTDEDD |
| ProTαC (56C/110C labeled) | GPSDAAVDTSSEITTKDLKEKKEVVEEAENGRDAPANGNAENEENGEQEA DNEVDEE**C**EEGGEEEEEEEEGDGEEEDGDEDEEAESATGKRAAEDDEDD DVDTKKQKTDED**C** |
| H1 (unlabeled) | TENSTSAPAAKPKRAKASKKSTDHPKYSDMIVAAIQAEKNRAGSSRQSIQK YIKSHYKVGENADSQIKLSIKRLVTTGVLKQTKGVGASGSFRLAKSDEPKKS VAFKKTKKEIKKVATPKKASKPKKAASKAPTKKPKATPVKKAKKKLAATPKK AKKPKTVKAKPVKASKPKKAKPVKPKAKSSAKRAGKKK |
| Protamine (unlabeled)* | MPRRRRSSSRPVRRRRRPRVSRRRRRRGGRRRR |
| Poly-L-lysine 50 | KKKKKKKKKKKKKKKKKKKKKKKKKKKKKKKKKKKKKKKKKKKKKKKKKK |
| Poly-L-arginine 50 | RRRRRRRRRRRRRRRRRRRRRRRRRRRRRRRRRRRRRRRRRRRRRRRRRR RRR |

**Table S1.** Amino acid sequences of polypeptides used in experiments and simulations. Cys residues introduced for labeling are indicated in bold. Unlabeled ProTα is a variant of human ProTα isoform 2, while ProTα 56C/110C is a variant of isoform 1 (64). The isoforms differ by a single Glu at position 39. *We note that the sequence given here and used in the simulations is a main component of the commercially available protamine we used (Sigma-Aldrich, product number P4020), but other components with similar lengths and amino acid composition are present in the preparation.

**Video S1 (separate file)**

**(Left)** All-atom explicit-solvent simulation of the ProTα–H1 condensate (total time 1 μs). One ProTα chain is highlighted in red (chain 60), and four interacting H1 chains are shown in different shades of blue. Other surrounding ProTα and H1 chains are shown semi-transparently in red and blue, respectively.
**(Right)** All-atom explicit-solvent simulation of the ProTα–protamine condensate (total time 1 μs). One ProTα chain is highlighted in red (chain 30), and six interacting protamine chains are shown in three different shades of purple. Other surrounding ProTα and protamine chains are shown semi-transparently in red and purple, respectively.

Both videos are centered on the center of mass of the highlighted ProTα chain. The video is shown at 2 ns per frame. To slightly smooth the motion, a filter with a time constant of 4 ns was applied to all frames. Protein hydrogen atoms, water molecules, and ions were omitted for clarity. (YouTube link to high-resolution version:
https://www.youtube.com/watch?v=E4Idah1J3N8&ab_channel=MilosIvanovic)

**Video S2 (separate file)**

**(Left)** All-atom explicit-solvent simulation of the ProTα–H1 condensate (total time 50 ns). One ProTα chain is highlighted in red (chain 59), and four interacting H1 chains are shown in different shades of blue. Other surrounding ProTα and H1 chains are shown semi-transparently in red and blue, respectively.

**(Right)** All-atom explicit-solvent simulation of the ProTα–protamine condensate (total time 50 ns). One ProTα chain is highlighted in red (chain 30), and six interacting protamine chains are shown in three different shades of purple. Other surrounding ProTα and protamine chains are shown semi transparently in red and purple, respectively.

Both videos are centered on the center of mass of the highlighted ProTα chain. The video is shown at 100 ps per frame. To slightly smooth the motion, a filter with a time constant of 200 ps was applied to all frames. Protein hydrogen atoms, water molecules, and ions were omitted for clarity. (YouTube link to high-resolution version:
https://www.youtube.com/watch?v=4G9GOYp-Fmw&ab_channel=MilosIvanovic)